\documentclass[12pt]{article}
\pdfoutput=1
\usepackage[english]{babel}
\usepackage[latin1]{inputenc}
\usepackage[normalem]{ulem}

\usepackage[text={16.4cm,23.7cm}]{geometry}
\usepackage{soul}
\usepackage{graphicx}
\usepackage{amsmath}
\usepackage{amssymb}
\usepackage{eurosym}
\usepackage{slashed}
\usepackage{hyperref}
\usepackage{wasysym}
\usepackage{feynmp}

\newcommand{\sub}[2]{#1_{\mathrm{#2}}}
\newcommand{\GeV}{~\mathrm{GeV}}
\newcommand{\TeV}{~\mathrm{TeV}}
\newcommand{\cm}{~\mathrm{cm}}

\begin{document}

\title{\vspace{-2cm} 
{\normalsize
\flushright TUM-HEP 912/13\\}
\vspace{0.6cm} 
\bf  High-energy neutrino signals from the Sun \\ 
in dark matter scenarios \\ with internal bremsstrahlung \\[8mm]}

\author{Alejandro Ibarra, Maximilian Totzauer and Sebastian Wild\\[2mm]
{\normalsize\it Physik-Department T30d, Technische Universit\"at M\"unchen,}\\[-0.05cm]
{\it\normalsize James-Franck-Stra\ss{}e, 85748 Garching, Germany}
}

\maketitle

\begin{abstract}
We investigate the prospects to observe a high energy neutrino signal from dark matter annihilations in the Sun in scenarios where the dark matter is a Majorana fermion that couples to a quark and a colored scalar via a Yukawa coupling. In this minimal scenario, the dark matter capture and annihilation in the Sun can be studied in a single framework. We find that, for small and moderate mass splitting between the dark matter and the colored scalar, the two-to-three annihilation $q \bar q g$ plays a central role in the calculation of the number of captured dark matter particles. On the other hand, the two-to-three annihilation into $q \bar q Z$ gives, despite its small branching fraction, the largest contribution to the neutrino flux at the Earth at the highest energies. We calculate the limits on the model parameters using IceCube observations of the Sun and we discuss their interplay with the requirement of equilibrium of captures and annihilations in the Sun and with the requirement of thermal dark matter production. We also compare the limits from IceCube to the limits from direct detection, antiproton measurements and collider searches.
\end{abstract}

\section{Introduction}
\label{sec:introduction}

The search for the particles which are presumed to be produced by dark matter annihilations is one of the primary goals in Astroparticle Physics (see \cite{Bertone:2004pz,Bergstrom:2000pn} for reviews on particle dark matter). This search is however hindered by the low fluxes expected from annihilations and by the existence of large, and still poorly understood, astrophysical backgrounds. Therefore, in order to detect a signal from annihilations, it is necessary to devise search strategies that mildly rely on the modelling of the backgrounds. Up to now, the most promising strategies to identify a signal from dark matter annihilations consist in {\it i)} the search for features in the measured energy spectrum of the products of annihilation and {\it ii)} the search for an excess of events following a spatial distribution with the morphology expected from an annihilation signal. 

Both strategies have been successfully applied to the search for annihilation signals in gamma-rays~\cite{Bringmann:2012ez}. Several spectral features have been identified~\cite{Srednicki:1985sf,Rudaz:1986db,Bergstrom:1988fp,Bergstrom:1989jr,Flores:1989ru,Bringmann:2007nk,Ibarra:2012dw} and searched for in the gamma-ray sky~\cite{Fermi-LAT:2013uma,Abramowski:2013ax,Bringmann:2012vr,Ibarra:2013eda}. Besides, the characteristic morphology of the signal expected from dark matter annihilations has been searched for in the Galactic Center region~\cite{Ackermann:2012qk,Abramowski:2011hc}, galaxy clusters~\cite{Ackermann:2010rg,Abramowski:2012au} and, notably, dwarf galaxies~\cite{Ackermann:2013yva,GeringerSameth:2011iw,Aleksic:2011jx,Abramowski:2010aa}. Both search strategies already lead to very stringent limits on the annihilation cross section which rule out some models. 
Pursuing these two promising strategies is, however, challenging when searching for antimatter particles produced in dark matter annihilations, since the propagation of the charged cosmic rays in the galaxy practically erases any spectral feature at production and any directional information, yielding a rather featureless spectrum and a fairly isotropic distribution of events. 
As a result, the limits from antimatter searches are typically weaker than in gamma-rays (see, however, the recent analyses \cite{Bergstrom:2013jra,Ibarra:2013zia}). Lastly, the search for dark matter annihilation using neutrinos as messengers shares many similarities with gamma-rays: both are electrically neutral particles and therefore the propagation in the galaxy does not significantly distort the energy spectrum nor the direction of these particles. Nevertheless, searches with neutrinos as messengers are limited by the small neutrino detection rate and by the poor energy and angular resolution of neutrino telescopes, thus making the search for features in the neutrino spectrum challenging. 
Indeed, the current limits on the annihilation cross section for annihilations into $\nu\bar\nu$~\cite{Abbasi:2012ws,Aartsen:2013mla} are significantly weaker than the limits for annihilations into $\gamma\gamma$. Interestingly, if dark matter particles interact with nucleons, they can be captured inside the Sun \cite{Press:1985ug} and subsequently annihilate, producing a high-energy neutrino flux in the direction of the center of the Sun~\cite{Silk:1985ax,Gould:1987ir}. The very characteristic spatial morphology of the signal then opens the possibility of probing the scattering cross section of dark matter particles with nucleons.  Conversely, the observation of an excess of high-energy neutrinos in the (time-dependent) direction of the Sun would be a clear signal of dark matter annihilations. The same phenomenon of capture might also occur in the Earth and other celestial bodies. 

In this paper we will concentrate in the possibility of detecting a high-energy neutrino flux originated in the annihilations of dark matter particles that have been previously captured inside the Sun or inside the Earth. Most analyses postulate that the dark matter particles had been captured and that their population is constant today, due to the equilibration between the processes of capture and annihilation. In this way, it is possible to set strong limits on the capture cross section from the non-observation of a neutrino flux originated in annihilations into, for example, $\nu\bar\nu$ or $W^+W^-$~\cite{Aartsen:2012kia}. However, this requires that the dark matter particle must couple to neutrinos or to weak gauge bosons, but also sizably to quarks to allow their capture. This scenario can be justified in concrete realizations although does not necessarily hold in general. 

We will then discuss the capture and the annihilation in a single Particle Physics framework. More concretely, we will consider a toy model where the dark matter particle is a Majorana fermion that couples to a quark and a scalar via a Yukawa coupling. The coupling to the quark leads to the capture of dark matter particles inside the Sun or the Earth. On the other hand, and as is well known, in this toy model the annihilation rate into a quark-antiquark pair is helicity- and velocity- suppressed. Hence, in the equilibration the internal bremsstrahlung processes ${\rm DM}\,{\rm DM}\rightarrow q \bar q V$, with $V$ a gauge boson, play a major role and, in fact, determine the minimal value of the coupling that allows equilibration inside the Sun or the Earth. Moreover, these processes determine the energy spectrum of neutrinos produced in the annihilation and accordingly the detection rate at the Earth (neutrino signals from two-to-three annihilations in leptophilic dark matter models have been considered previously in \cite{Bell:2012dk,Fukushima:2012sp}). The parameter space of this toy model is further constrained by the requirement of reproducing the correct dark matter abundance from thermal production, as well as from direct, indirect and collider dark matter searches. Concretely, we will consider, following \cite{Garny:2012eb,Garny:2013ama}, the limits from the XENON100 experiment \cite{Aprile:2012nq}, the PAMELA measurements on the antiproton-to-proton fraction \cite{Adriani:2010rc}, as well as collider limits from the CMS experiment~\cite{Chatrchyan:2013lya}. All these limits, together with the requirement of building up a population of dark matter particles inside the Sun or the Earth, will then allow us to assess the prospects to observe a high energy neutrino signal from the Sun or the Earth in neutrino telescopes, concretely in IceCube.

The paper is organized as follows. In Section \ref{sec:toy_model} we present the toy model under consideration as well as the properties relevant for our analysis. Then, in Section \ref{sec:capture} we describe the process of capture of dark matter particles in the Sun and in the Earth and we present the constraints on the parameter space of our toy model from the requirement of equilibration between capture and annihilation. In Section \ref{sec:HE_flux} we calculate the neutrino flux from the Sun, emphasizing the role of the internal bremsstrahlung process in generating a high-energy component. 
In Section \ref{sec:limits} we present the limits on the parameter space which stem from the negative searches of a high-energy neutrino excess in the direction of the Sun in IceCube and we compare our limits to those from other experiments. Lastly, in Section \ref{sec:conclusions} we present our conclusions.

\section{Toy model with internal bremsstrahlung}
\label{sec:toy_model}

We consider a toy model where the dark matter particle is a Majorana fermion $\chi$, singlet under the Standard Model gauge group, which interacts with a quark and a scalar $\eta$ via a Yukawa interaction with coupling constant $f$. The Lagrangian of the model is given by:
\begin{align}
  {\cal L}={\cal L}_{\rm SM}+{\cal L}_{\chi}+{\cal L}_\eta+ {\cal L}_{\rm int}
  \;.
\end{align} 
Here, ${\cal L}_{\rm SM}$ is the Standard Model Lagrangian, while ${\cal L}_{\chi}$ and ${\cal L}_\eta$ are the parts of the Lagrangian involving just the new fields $\chi$ and $\eta$ and which are given, respectively, by:
\begin{align}
 \begin{split}
    {\cal L}_\chi&=\frac12 \bar \chi^c i\slashed {\partial} \chi
    -\frac{1}{2}m_\chi \bar \chi^c\chi\;, \; \text{and}\\ {\cal L}_\eta&=(D_\mu
    \eta)^\dagger  (D^\mu \eta)-m_\eta^2 \eta^\dagger\eta -\frac{1}{2}\lambda_2 (\eta^\dagger \eta)^2\;,
  \end{split}
\end{align}
Here $D_\mu$ denotes the usual covariant derivative. Lastly, the interaction term in the Lagrangian is given by
\begin{align}
  {\cal L}_{\rm int} &= -\lambda_3(\Phi^\dagger \Phi)(\eta^\dagger \eta) - f \bar \chi q_R \eta+{\rm h.c.} \;,
\label{eq:singlet-qR}
\end{align}
where $\Phi$ is the Standard Model Higgs boson. Here we have assumed for concreteness that the quark is right-handed, hence the quantum numbers of the scalar under $SU(3)_c\times SU(2)_L\times U(1)_Y$ are $\left(\bar{3},1, q_f \right)$,  where $q_f$ is the right-handed quark electric charge (which in this case coincides with minus the hypercharge).

This toy model has been thoroughly discussed in relation to indirect dark matter searches. The lowest order annihilation process $\chi\chi\to q\bar q$ is helicity- and velocity- suppressed, resulting in an annihilation cross section which is far below the reach of present and foreseeable indirect dark matter search experiments. Interestingly, the helicity suppression is lifted by the associated emission of a spin-one boson~\cite{Bergstrom:1989jr,Flores:1989ru}.\footnote{The emission of a Higgs boson also lifts the helicity suppression, as recently pointed out in \cite{Luo:2013bua}. The cross section, however, strongly depends on the parameters of the scalar potential, therefore we will neglect these processes in our analysis for simplicity.}  More specifically, the annihilation processes $\chi\chi\to q\bar q V$, with $V$ being a gluon~\cite{Flores:1989ru,Drees:1993bh,Barger:2006gw}, a $Z/W$-boson~\cite{Garny:2011ii,Garny:2011cj,Ciafaloni:2011sa,Ciafaloni:2011gv,Ciafaloni:2012gs,Bell:2011if,Bell:2011eu} (see also \cite{Kachelriess:2009zy,Ciafaloni:2010ti}) or a photon~\cite{Bergstrom:1989jr,Flores:1989ru,Bringmann:2007nk} can dominate over the two-to-two process, provided the scalar that mediates the interaction is close in mass to the dark matter particle~\cite{Garny:2011cj}. Moreover, in this regime a sharp spectral feature arises in the gamma-ray spectrum which, if observed, would constitute a smoking gun for dark matter detection. \footnote{Sharp spectral features from internal bremsstrahlung also arise in the annihilation of scalar dark matter particles \cite{Toma:2013bka,Giacchino:2013bta,Garcia-Cely:2013zga}.} Therefore, new opportunities open for indirect dark matter searches using antimatter, gamma-rays and, as we will discuss in this paper, neutrinos. 

\begin{figure}
\begin{center}
\includegraphics[width=0.49\textwidth]{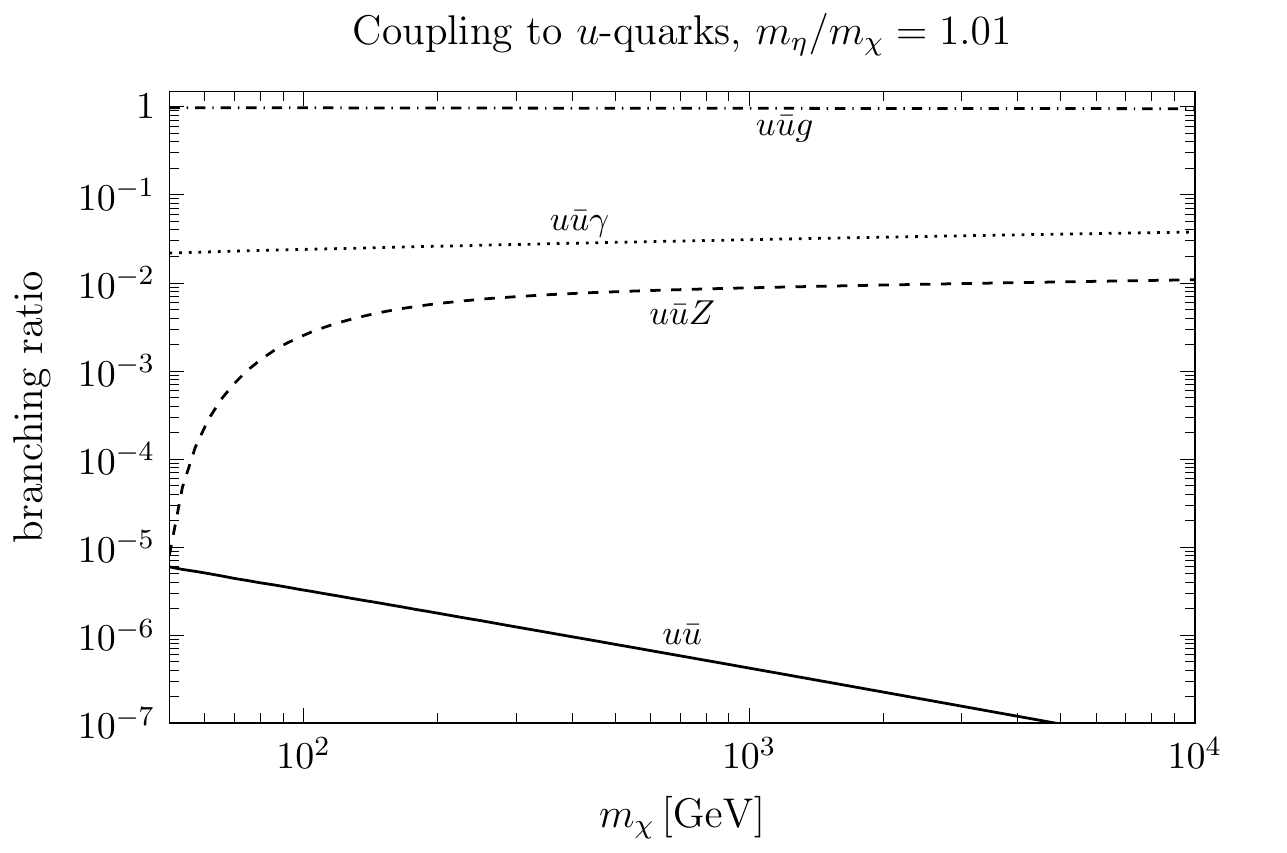}
\includegraphics[width=0.49\textwidth]{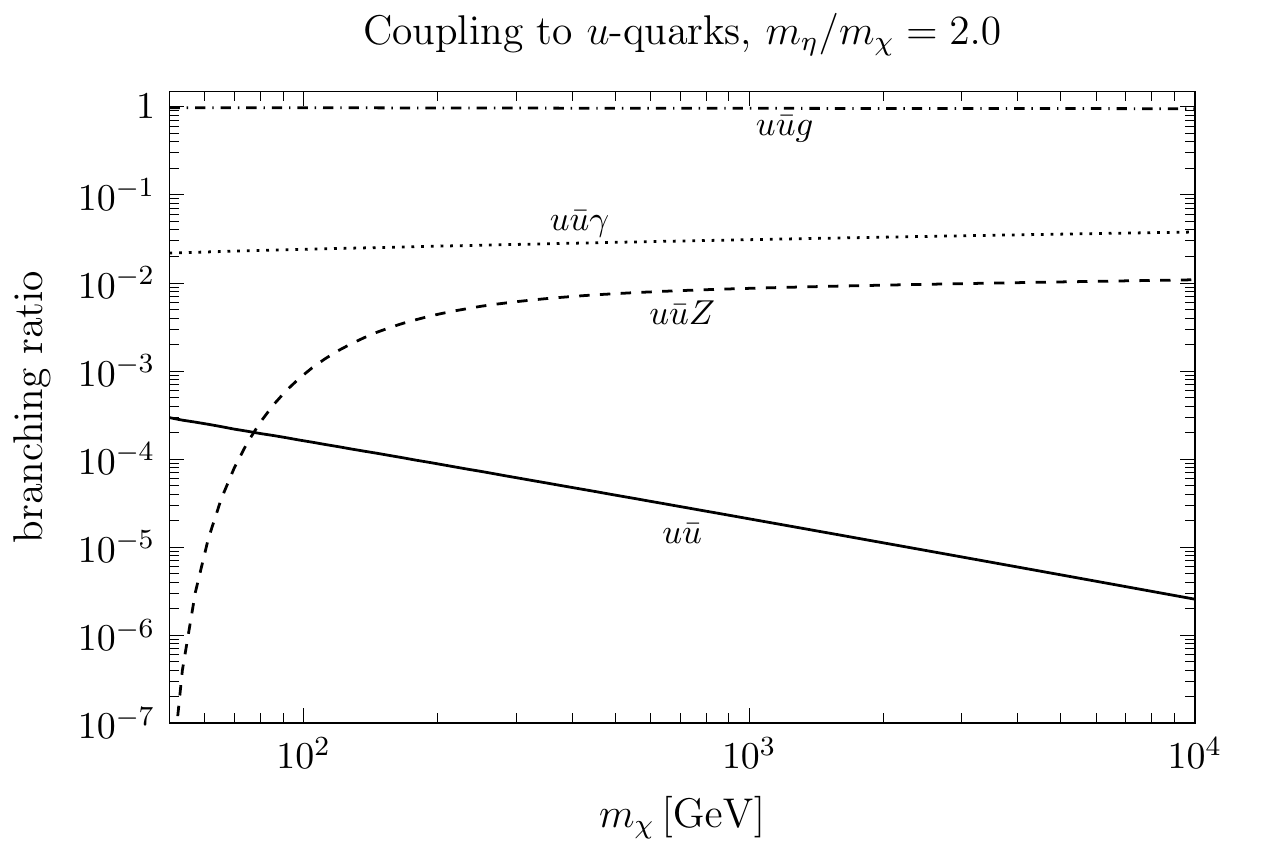}\\
\includegraphics[width=0.49\textwidth]{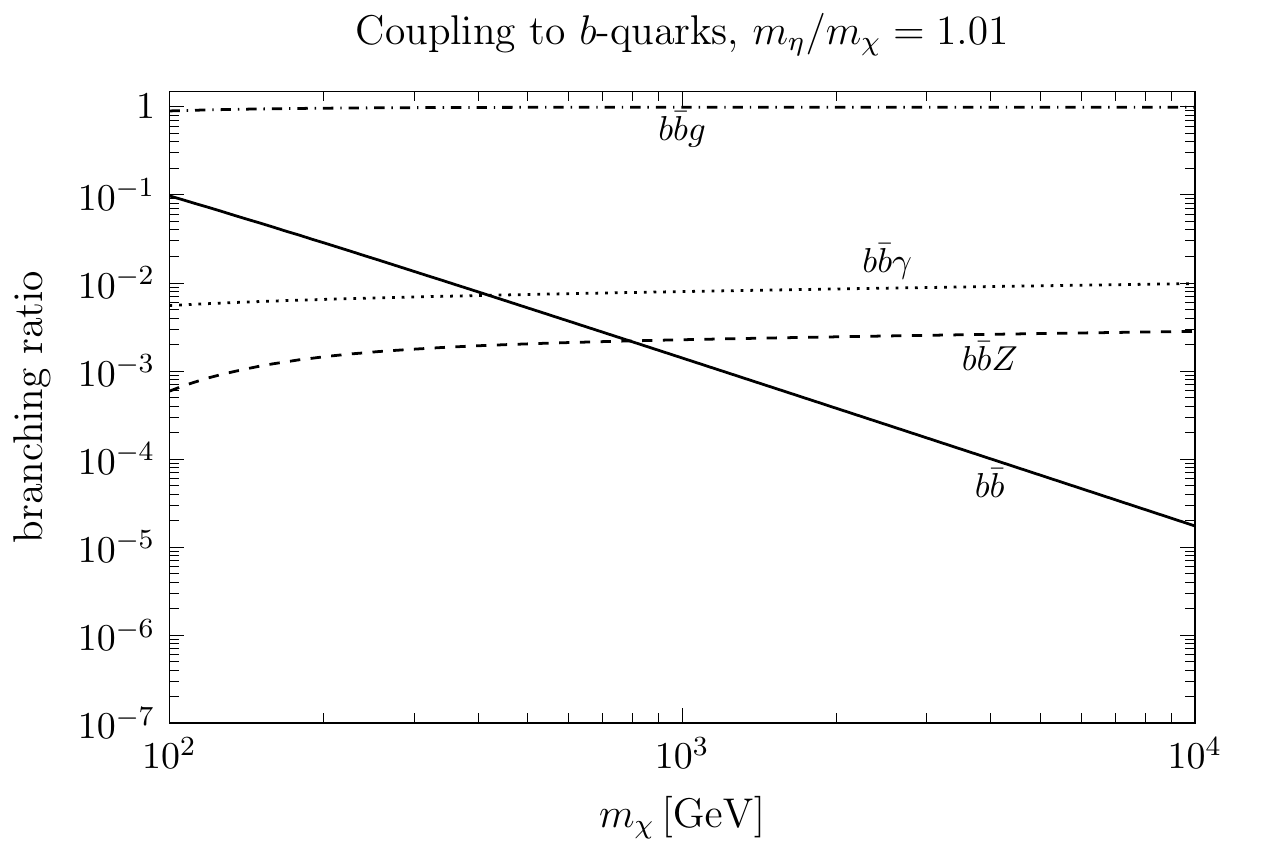}
\includegraphics[width=0.49\textwidth]{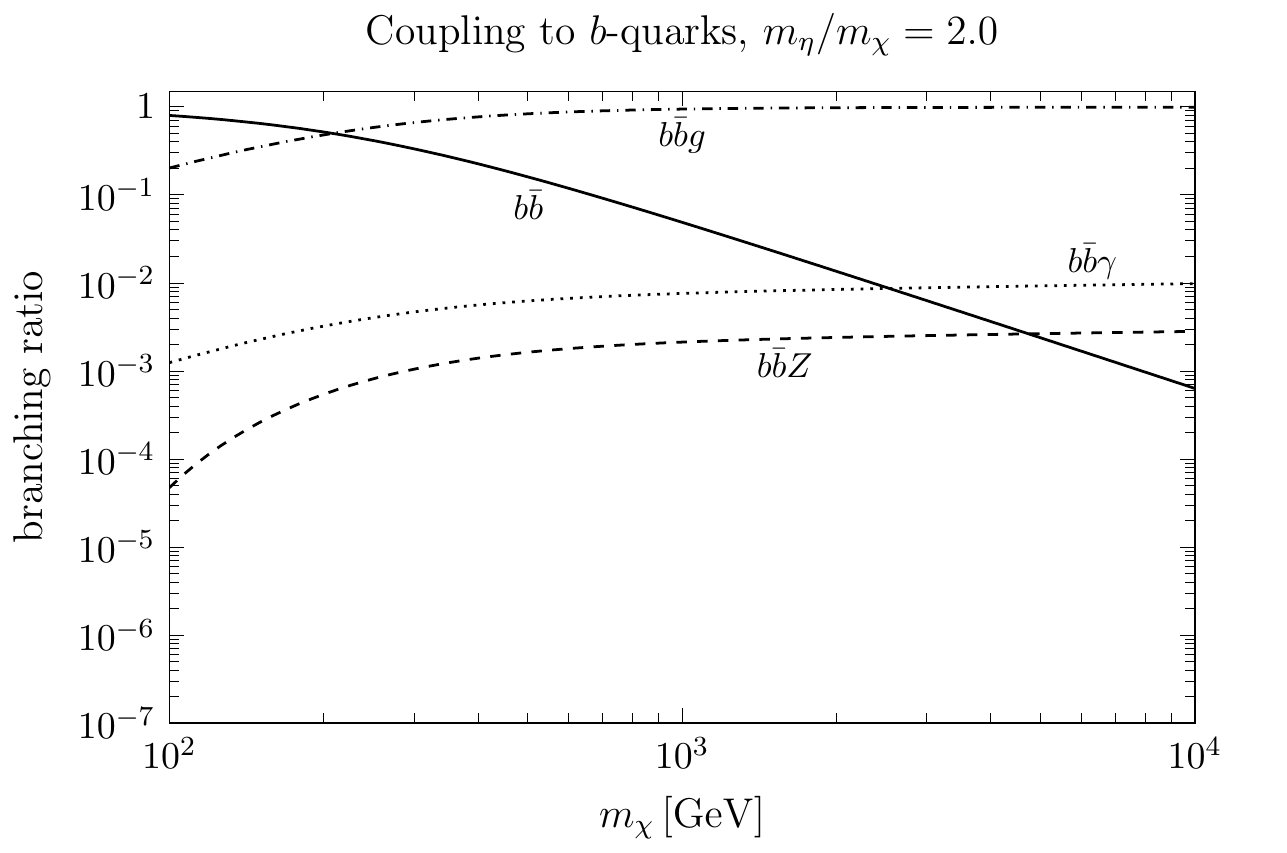}
\caption{Branching fraction of the annihilation $\chi \chi \rightarrow q \bar{q} (V)$, with $V$ a gauge boson and $q$ a right-handed up-quark (upper panels) or bottom-quark (lower panels), for the mass ratios $m_\eta/m_\chi=1.01$ (left panels) and 2.0 (right panels).} 
\label{fig:branchingratios}
\end{center}
\end{figure}

The relative importance of the various annihilation channels is illustrated in Fig. \ref{fig:branchingratios}, where we show the branching ratios as a function of the dark matter mass for the exemplary cases $m_\eta/m_\chi=1.01$, namely a scenario with large degeneracy, and $m_\eta/m_\chi=2$. In these plots, and in the rest of the paper, we neglect the quark masses in the calculation of the rates of the two-to-three processes. It is worth recalling that when $m_\chi\gg M_Z$, which is the range of masses relevant for neutrino telescopes, the ratios of cross sections of the two-to-three processes are fairly independent of the mass splitting and take the values~\cite{Garny:2011ii}:
\begin{equation}
\begin{array}{ccccccc}
  \sigma v(\chi\chi\to g u_R\bar u_R) & : & \sigma v(\chi\chi\to \gamma u_R\bar u_R) & = & 3\alpha_s(m_{\rm DM})/\alpha_{em} & \simeq & 32.2\;, \\
  \sigma v(\chi\chi\to Z u_R\bar u_R) & : & \sigma v(\chi\chi\to \gamma u_R\bar u_R) & = & \tan^2(\theta_W) & = & 0.30\;,
\label{eq:ratios-uR}
\end{array}
\end{equation}
for couplings to right-handed up-type quarks and 
\begin{equation}
\begin{array}{ccccccc}
  \sigma v(\chi\chi\to g d_R\bar d_R) & : & \sigma v(\chi\chi\to \gamma d_R\bar d_R) & = & 12\alpha_s(m_\chi)/\alpha_{em} & \simeq & 129\;, \\
  \sigma v(\chi\chi\to Z d_R\bar d_R) & : & \sigma v(\chi\chi\to \gamma d_R\bar d_R) & = & \tan^2(\theta_W) & = & 0.30\;.
\label{eq:ratios-dR}
\end{array}
\end{equation}
for right-handed down-type quarks. The numerical values in these formulas have been obtained evaluating, for illustration, the strong coupling constant at the scale $m_\chi=1\TeV$.

The dark matter interaction to a quark in Eq.~(\ref{eq:singlet-qR}) not only induces dark matter annihilations but also the elastic scattering of dark matter particles with a nucleus in a detector, thus opening the possibility of observing signals also in direct dark matter search experiments and, as we will see in Section \ref{sec:capture}, of capturing dark matter particles inside the Sun or the Earth. The scattering cross sections for this model were calculated in  \cite{Hisano:2010ct,Drees:1993bu,Ellis:2008hf}. The spin dependent part reads
 \begin{align}
   \sub{\sigma}{SD}^p = \frac{12 \mu_p ^2}{\pi} \left(\Delta q\right)^2 \frac{f^4}{64 \left[ m_\eta^2 -\left( m_\chi+ m_q\right)^2\right]^2} \,,
   \label{eq:Definition_sigmaSD}
  \end{align}
where $\mu_p$ is the reduced mass of the proton-dark matter system and $\Delta q$ parametrizes the quark spin content of the proton for the light quark  to which the dark matter couples; for heavy quarks, $\Delta q=0$. Besides, the spin independent cross section with protons reads:
 \begin{align}
   \sub{\sigma}{SI}^{p} &= \frac{4 \mu_p ^2}{\pi} f_p^2 \,,
   \label{eq:Definition_sigmaSI}
  \end{align}
  where
  \begin{align}
    f_{p} &= m_{p} \left( 
    f_{T_q}^{p} \frac{m_\chi}{2} \frac{f^2}{8 \left[ m_\eta^2 -\left( m_\chi+ m_q\right)^2\right]^2}  +  \frac{3}{2} m_\chi \frac{f^2}{8 \left[ m_\eta^2 -\left( m_\chi+ m_q\right)^2\right]^2}  \left[ q^{p}\left(2\right) + \bar{q}^{p} \left(2\right) \right] - \frac{8}{9} \pi b  f_{T_g}^{p}
    \right) \,,
    \label{eq:Definition_fp}
  \end{align}
with an analogous expression for the neutron-dark matter cross section (note that we do not assume isospin invariance). Here, $b$ is a function of the masses of the respective quark, the dark matter particle and the scalar mediator $\eta$ which we take from \cite{Drees:1993bu,Jungman:1995df}. Moreover, $f_{T_q}^{p/n}$ ($f_{T_g}^{p/n}$) parametrizes the scalar quark (gluon) current in the proton/neutron; we use the values and uncertainties from \cite{Ellis:2008hf}. Lastly, $q^{p/n} \left(2\right)$ are the second moments of the particle distribution function of the quark inside the nucleon, given in \cite{Hisano:2010ct}. Note that both the spin independent and the spin dependent cross sections get enhanced when the dark matter particle and the intermediate scalar $\eta$ are close in mass, which is precisely the region of the parameter space relevant for indirect dark matter searches in this scenario \cite{Garny:2012eb}. The expressions for the cross sections, Eqs.~(\ref{eq:Definition_sigmaSD},\ref{eq:Definition_sigmaSI}), on the other hand, become divergent when $m_{\eta} = m_{\chi} + m_q$ and therefore it is necessary to include close to the resonance the width of the exchanged scalar. However, and in order to avoid the model dependence introduced by the modelling of the scalar decay, we will restrict our analysis to values of the model parameters sufficiently away from the resonance, concretely we will only consider $m_\eta-m_\chi\geq 2 m_q$.

\section{Dark matter capture in the Sun and in the Earth}
\label{sec:capture}

In our toy model we have postulated that the dark matter particle couples to a quark, therefore, a dark matter particle traversing the Sun could scatter-off a nucleus in the solar interior and lose energy. After subsequent scatterings, the dark matter particles eventually sink to the solar core where they accumulate. At the same time, the dark matter population in the solar core is depleted by annihilations and -- in case of light dark matter particles -- by thermal evaporation. The time evolution of the number of dark matter particles $N$ in the solar core is then described by the following differential equation~\cite{Griest:1986yu}:
    \begin{align}
        \frac{dN}{dt} = \sub{\Gamma}{C} - \sub{C}{A} N^2 - \sub{C}{E} N \, , \label{eq:DEDMDensitySun} 
    \end{align}
where $\sub{\Gamma}{C}$ is the capture rate, $\sub{C}{A}$ the annihilation constant and $\sub{C}{E}$ the evaporation constant. 

For dark matter masses above $\sim 10\GeV$ the evaporation term can be safely neglected \cite{Jungman:1995df,Busoni:2013kaa}. Then, after solving Eq.\eqref{eq:DEDMDensitySun}, one finds an annihilation rate as a function of time given by
  \begin{align}
      \sub{\Gamma}{A}(t) &=\frac{1}{2} \sub{C}{A} N(t)^2= \frac{1}{2} \sub{\Gamma}{C} \tanh^2 \left( t / \tau \right) \, , \label{eq:SolutionDMDensitySun}
  \end{align}
where 
  \begin{align}
       \tau &= \frac{1}{\sqrt{\sub{\Gamma}{C} \sub{C}{A}}} \, . \label{eq:EquilibrationTime} 
  \end{align}
The annihilation rate reaches a maximum when $t\gg \tau$. In this regime, captures and annihilations are in equilibrium and the annihilation rate is determined only by the capture rate: $\sub{\Gamma}{A}= \sub{\Gamma}{C}/2$. It is important to note that, despite the Earth and the Sun have been capturing dark matter particles since their formation approximately $4.5\times 10^9 \mathrm{y}$ ago, for some choices of the model parameters equilibrium might have not been reached. Whether dark matter particles are or not in equilibrium in the Sun or the Earth is determined by the values of the capture rate and the annihilation constant, {\it cf.} Eq. \eqref{eq:EquilibrationTime}. The capture rate can be calculated from the scattering cross section of the process dark matter-nucleon and relies on assumptions on the density and velocity distributions of dark matter particles in the Solar System, as well as on the composition and density distribution of the interior of the Sun or the Earth. In our analysis we will use DarkSUSY to evaluate the capture rate from the scattering cross sections in Eqs.~(\ref{eq:Definition_sigmaSD},\ref{eq:Definition_sigmaSI}), adopting for concreteness a local dark matter density $\sub{\rho}{local} = 0.4 \GeV/\cm^3$ \cite{Catena:2009mf} and a homogeneous Maxwell-Boltzmann dark matter distribution with velocity dispersion $v_0 = 230\pm 30~\mathrm{km}/\mathrm{s}$ \cite{Pato:2010zk}. On the other hand, the annihilation constant for the Sun is approximately given by \cite{Jungman:1995df}
  \begin{align}
   \sub{C}{A}^{\mathrm{Sun}} = 1.6 \times 10^{-52}\, \mathrm{s^{-1}}\,
   \left(\frac{\langle \sigma_A v\rangle}{3 \times 10^{-26} \, \mathrm{cm^3s^{-1}}}\right)\, 
   \left(\frac{m_\chi}{\mathrm{TeV}}\right)^{3/2} \,,
   \label{eq:Definition_CA}
  \end{align}
while for the Earth by
  \begin{align}
  \sub{C}{A}^{\mathrm{Earth}} = 5.3 \times 10^{-49}\, \mathrm{s^{-1}}\,
  \left(\frac{\langle \sigma_A v\rangle}{3 \times 10^{-26} \, \mathrm{cm^3s^{-1}}}\right)\, 
  \left(\frac{m_\chi}{\mathrm{TeV}}\right)^{3/2} \,,
   \label{eq:Definition_CA_Earth}
  \end{align}
where the annihilation cross sections for $\chi \chi \rightarrow q \bar{q}(V)$ are given in \cite{Garny:2011ii}.

Both the capture rate and the annihilation constant depend on the fourth power of the Yukawa coupling $f$, thus allowing to define a lower limit on the coupling constant $f$ from requiring equilibration. Taking for concreteness the case of the Sun, the equilibration condition reads  $t_\odot\simeq 1/\sqrt{\Gamma_C C_A}$, where $t_\odot \approx 4.5\times 10^9 \mathrm{y}$. More specifically, we define the equilibrium coupling constant $f_{eq}$ from the requirement that the argument of the hyperbolic tangent in Eq.\eqref{eq:SolutionDMDensitySun} is equal to 1 for $t=t_\odot$, namely 
 \begin{align}
    t_{\odot} \sqrt{\sub{\Gamma}{C}^{f=1} \sub{C}{A}^{f=1}} \sub{f}{eq}^4 \overset{!}{=} 1 \,, \label{eq:DefinitionMinimalCouplingEquilibrium}
  \end{align}
 where $\sub{\Gamma}{C}^{f=1}$ and $\sub{C}{A}^{f=1}$ are, respectively, the capture  and annihilation constants for $f=1$. This implies that the ratio of annihilation rate and capture rate can be cast as:
  \begin{align}
   \frac{2 \sub{\Gamma}{A}}{\sub{\Gamma}{C}} = \tanh^2 \left[\left(\frac{f}{\sub{f}{eq}} \right)^4 \right] \, . \label{eq:EquilibriumMeasure}
  \end{align}
It follows from this equation that for values of the coupling larger than the equilibrium value, the annihilation rate, and hence the signal strength, reaches exponentially fast the maximal value. On the other hand, when the coupling is smaller than the equilibrium value, the annihilation rate exponentially decreases with $(f/f_{\rm eq})^4$. For example, for $f=f_{\rm eq}/2$, the signal strength drops to about 0.4\% of the maximal (equilibrium) signal strength. 

\begin{figure}
\begin{center}
\includegraphics[width=0.49\textwidth]{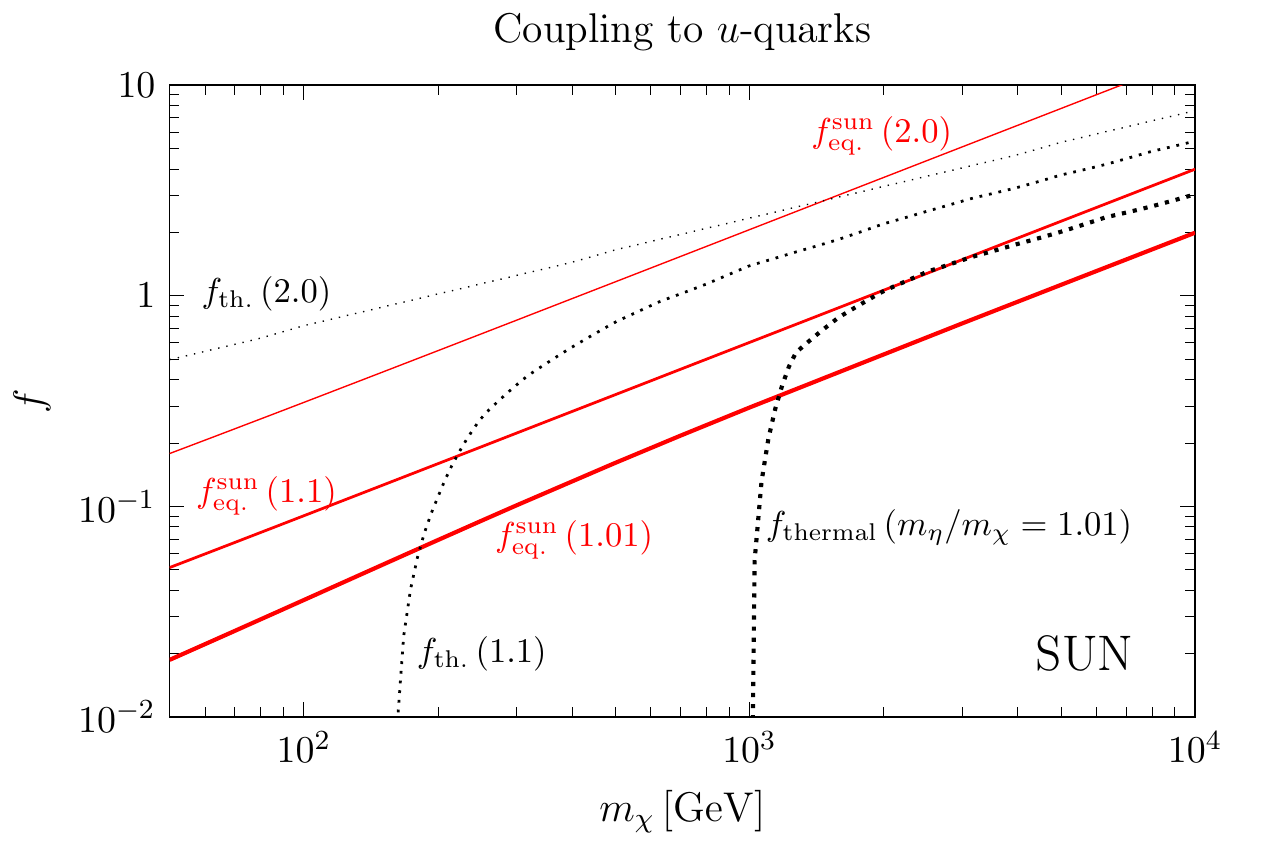}
\includegraphics[width=0.49\textwidth]{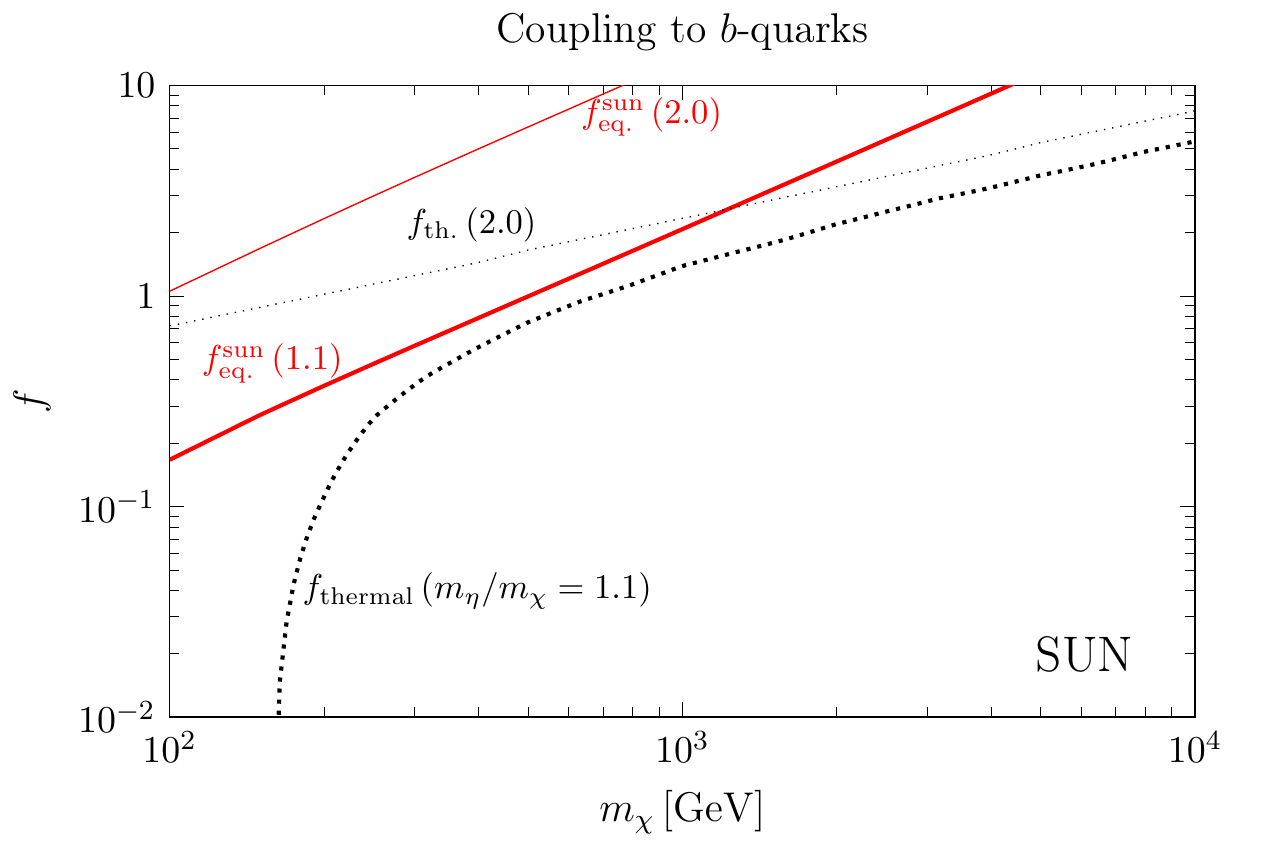}\\
\includegraphics[width=0.49\textwidth]{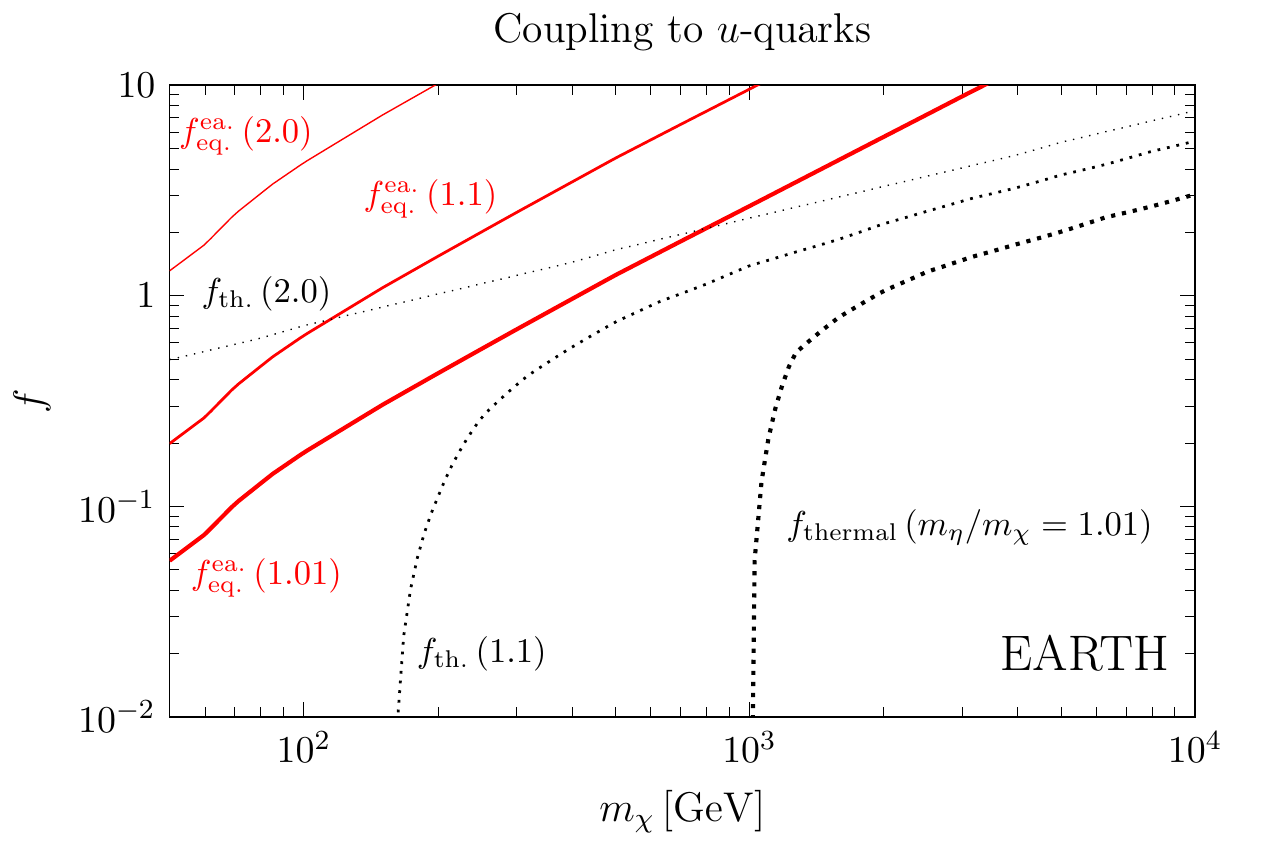}
\includegraphics[width=0.49\textwidth]{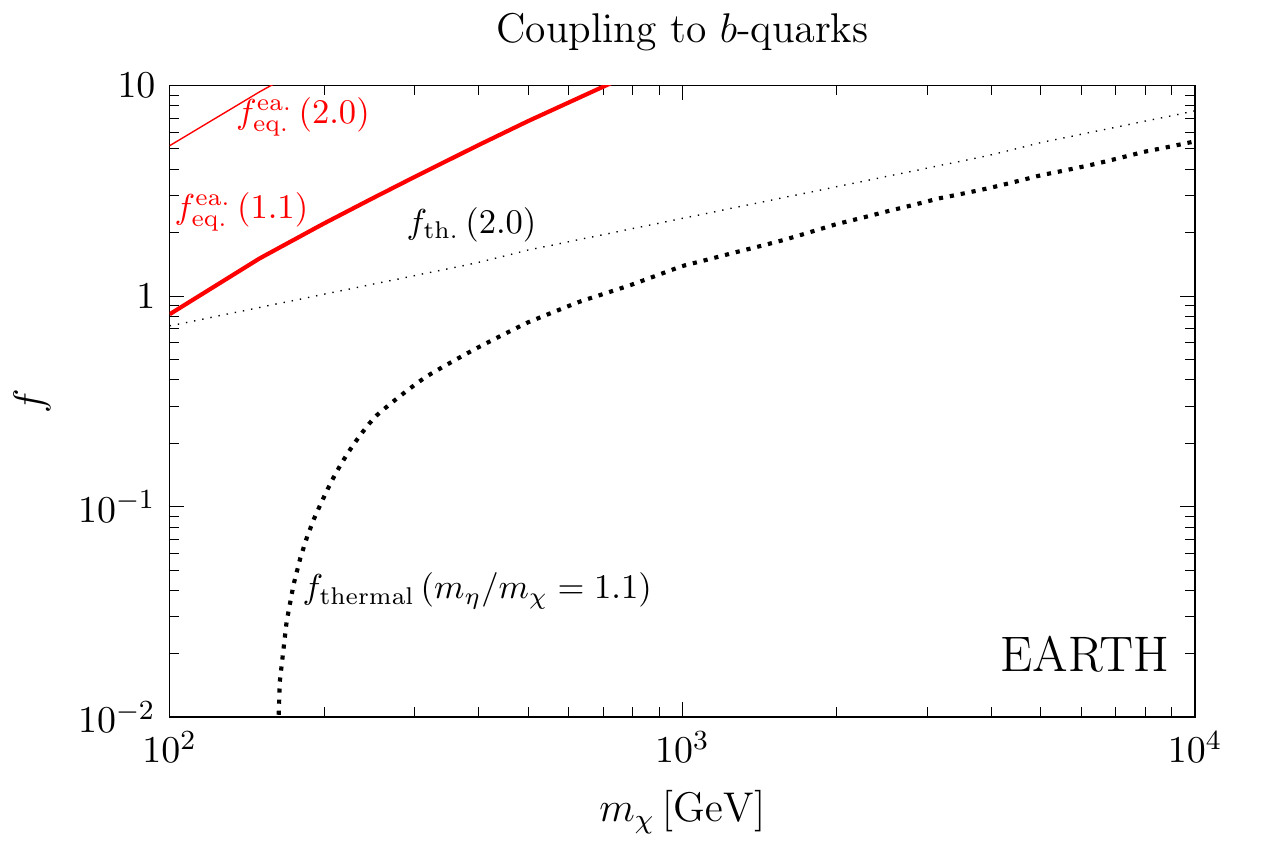}
\caption{Minimal values of the coupling $f$ for various mass ratios from the requirement of equilibration between dark matter captures and annihilations in the Sun (upper panels) and in the Earth (lower panels), for coupling to up-quarks (left panels) and to bottom-quarks (right panels). Also indicated as dashed lines are the values of the couplings that generate the observed dark matter abundance via thermal production. }
\label{fig:equilibrium}
\end{center}
\end{figure}

We show in Fig.~\ref{fig:equilibrium} as red solid lines the minimal values of the coupling constant which are required to achieve equilibration between capture and annihilation for dark matter particles that couple to the right-handed up- (left plot) or bottom-quarks (right plot) as a function of the dark matter mass, for the mass ratios $m_\eta/m_\chi=1.01$, 1.1 and 2.0. These lines scale with the mass as $m_\chi^{17/16}$ for the spin independent dominated capture (as is the case of couplings to the bottom quark) and as $m_\chi^{13/16}$ for the spin dependent dominated capture (as is the case of couplings to the up-quark).
\footnote{These scalings follow from the dependence of the annihilation cross section with the mass $\langle \sigma v \rangle \sim m_\chi^{-2}$, which implies an annihilation rate $C_A^{f=1}\sim m_\chi^{-1/2}$, as well as the dependences of the capture cross sections $\sigma^{SD}\sim m_\chi^{-4}$ and  $\sigma^{SI}\sim m_\chi^{-6}$, which translate into a capture rate $\Gamma^{f=1}_C\sim m_\chi^{-6}$ and  $\Gamma^{f=1}_C\sim m_\chi^{-8}$ for the spin dependent and the spin independent capture, respectively. Inserting these dependences in Eq.~(\ref{eq:DefinitionMinimalCouplingEquilibrium}) one obtains $f_{eq}\sim m_\chi^{13/16}$ and $\sim m_\chi^{17/16}$ for the spin dependent and the spin independent dominated captures.} Below these lines the annihilation rate becomes very suppressed, due to the small number of particles trapped inside the Sun, and therefore observing a high energy neutrino signal from the Sun becomes very challenging, if not impossible in practice.

We also show, for comparison, the values of the coupling $f$ calculated in \cite{Garny:2012eb}, that lead to the correct dark matter relic density. Note the existence of a lower limit on the dark matter mass from the requirement of reproducing the correct relic abundance when the scalar $\eta$ is highly degenerate in mass with the dark matter particle, and which is due to the impact of coannihilations in the mass degenerate regime (for details, see  \cite{Garny:2012eb},\cite{Garny:2013ama}). Concretely, for couplings to up-quarks this lower limit is roughly 1 TeV and 150 GeV for $m_\eta/m_\chi=1.01$ and 1.1, respectively.
Furthermore, for thermally produced dark matter particles, there are upper and lower limits on the dark matter mass from the requirement of equilibration between captures and annihilations in the Sun. Namely, $m_\chi\gtrsim 1\TeV$, $m_\chi\gtrsim 200\GeV$ and $m_\chi\lesssim 1\TeV$ for $m_\eta/m_\chi=1.01$, 1.1 and 2.0, respectively. For masses not fulfilling these conditions, the number of dark matter particles captured in the Sun is suppressed compared to the equilibrium values and, accordingly, the annihilation rate. Interestingly, for dark matter particles coupling only to bottom quarks, the coupling required to thermally produce the dark matter particles is smaller than the lower limit from the equilibration condition Eq.~(\ref{eq:DefinitionMinimalCouplingEquilibrium}) (this can be understood from the fact that $\sigma_{\text{SD}} = 0$). Hence considerably higher couplings are needed in order to achieve equilibrium and therefore in this scenario the commonly used approximation $\Gamma_A=\Gamma_C/2$ does not hold but instead the annihilation rate is always smaller than this value. 

The values of the coupling that lead to equilibration in the Earth are shown in Fig.~\ref{fig:equilibrium}, bottom plots. As apparent from the plots, within our framework a thermal relic can never be in equilibrium in the Earth. Therefore, the prospects for the detection of a neutrino flux from the center of the Earth are poor and hence we will restrict our discussion in the following to neutrino fluxes from dark matter annihilations in the Sun.

\section{High energy neutrino flux from the Sun} 
\label{sec:HE_flux}

The quarks and gauge bosons produced in the annihilations $\chi \chi \rightarrow q  \bar{q} (V)$ will hadronize and decay, eventually producing a neutrino flux. The dense medium inside the Sun where the annihilations take place significantly affects the energy spectrum of neutrinos, since the particles produced in the annihilations may lose a substantial fraction of their energy due to electromagnetic and strong interactions with the medium before decaying. We have simulated the energy loss of particles in the interior of the Sun following \cite{Ritz:1987mh}. We first calculate the primary spectrum of particles produced in the annihilation with PYTHIA8.1~\cite{Sjostrand:2007gs} interfaced with CalcHEP~\cite{Pukhov:1999gg, Pukhov:2004ca}. We then assume that the muons and light hadrons, such as pions and kaons, are stopped in the Sun before decaying and hence produce neutrinos with energies below 1 GeV, which we neglect in our analysis (the low energy neutrino flux from dark matter annihilations can also be used to probe dark matter models, as discussed in \cite{Rott:2012qb},\cite{Bernal:2012qh}). On the other hand, taus and heavy hadrons, such as charmed or beauty hadrons, decay in flight after losing a fraction of their kinetic energy. We have simulated the energy loss and subsequent decay in flight of these particles in an event-by-event basis following the chain of scatterings they undergo inside the Sun from the first scattering to the last (contrary to \cite{Blennow:2007tw} which follows the chain of scatterings from the last to the first; the difference between both approaches turns out to be, however, negligible for our analysis).
In addition to the neutrinos produced in hadronic and leptonic decays there is a component in the spectrum of hard neutrinos produced by the prompt decay of the $Z$ boson from the annihilation $\chi \chi \rightarrow q  \bar{q}  Z$. This process, despite having a small branching ratio, can be an important source of high energy neutrinos from the Sun and can even be the dominant contribution to the flux at the highest energies, as we will show below. 

Neutrinos produced in dark matter annihilations then propagate from the solar interior to the surface, undergoing flavour oscillations and scatterings off solar matter, and then from the solar surface to a neutrino telescope at the Earth, undergoing flavour oscillations in vacuum. To calculate the fluxes at the Earth, we use  WimpSim \cite{Blennow:2007tw}, a publicly available Monte Carlo code that simulates, in an event-by-event basis, the propagation of neutrinos from the production point in the Solar center to the Earth, including the effect of neutrino interactions in matter and three-flavour neutrino oscillations; for the calculation we have adopted the most recent best fit values for the neutrino oscillation parameters~\cite{Beringer:1900zz}. 

\begin{figure}
\begin{center}
\includegraphics[width=0.49\textwidth]{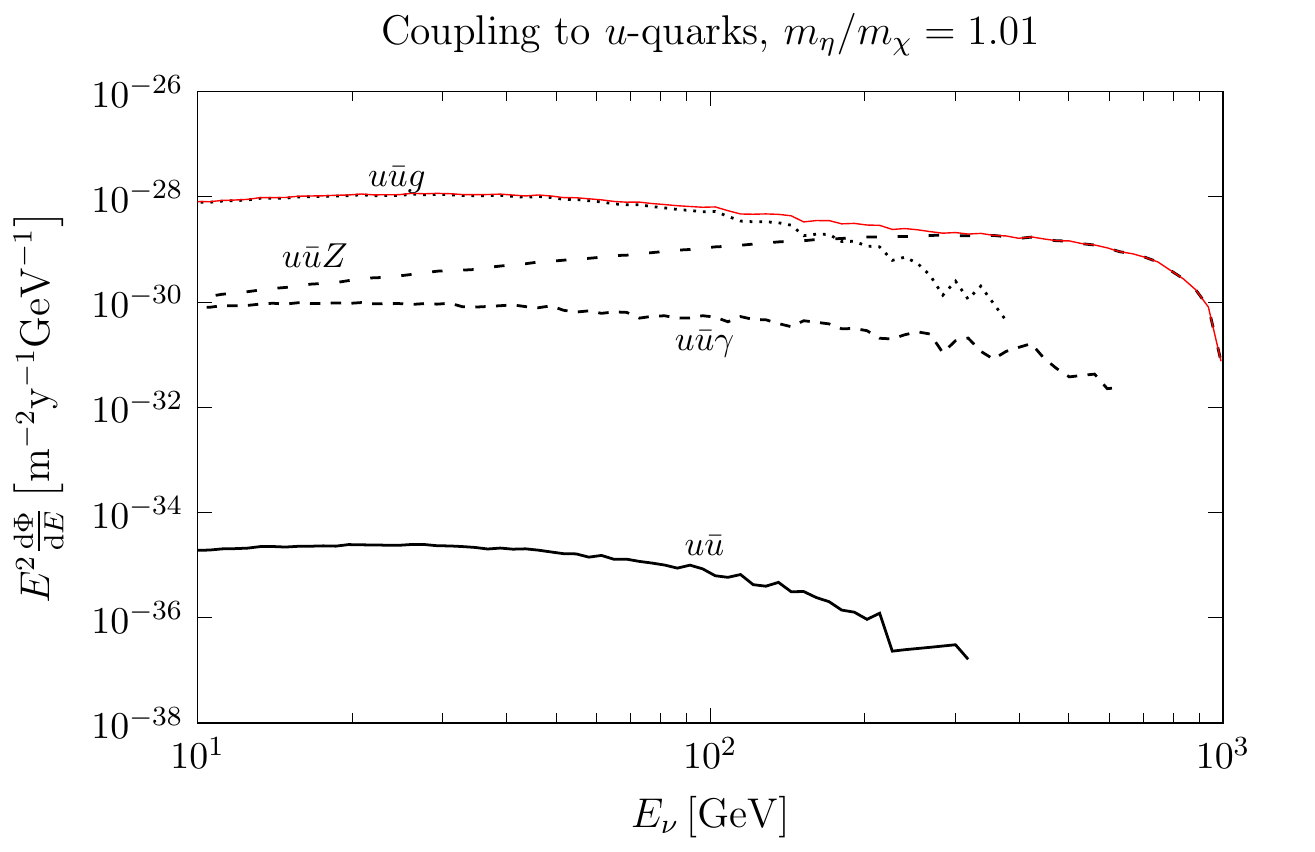}
\includegraphics[width=0.49\textwidth]{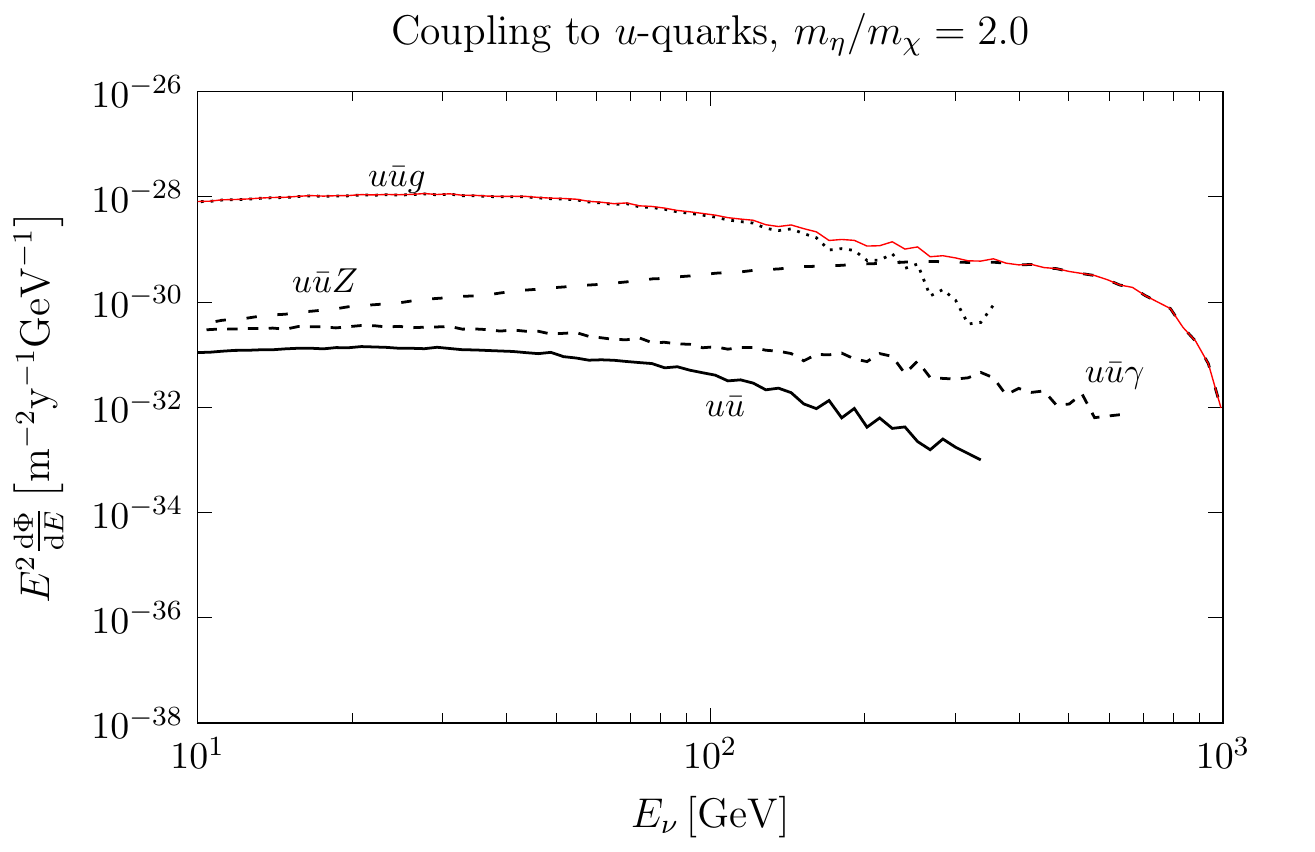}\\
\includegraphics[width=0.49\textwidth]{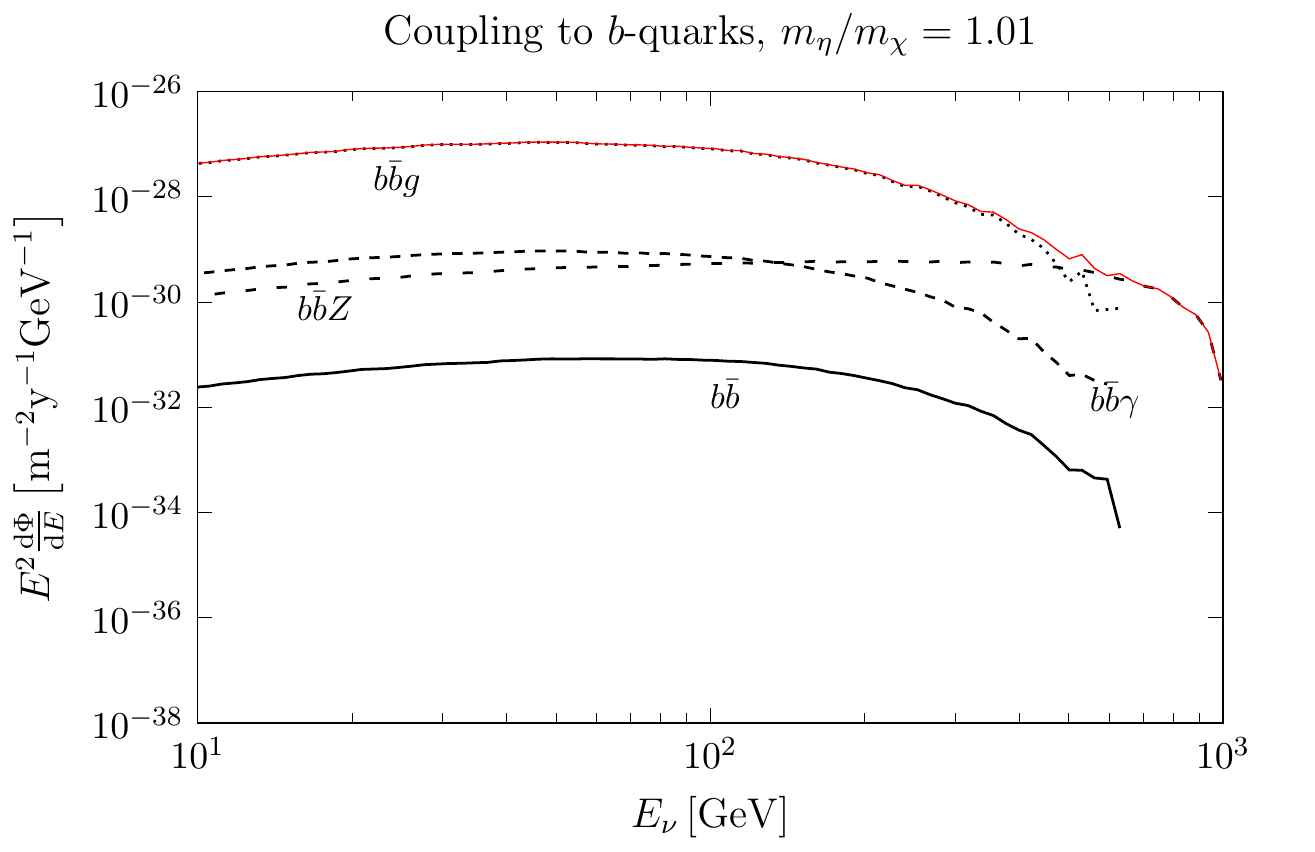}
\includegraphics[width=0.49\textwidth]{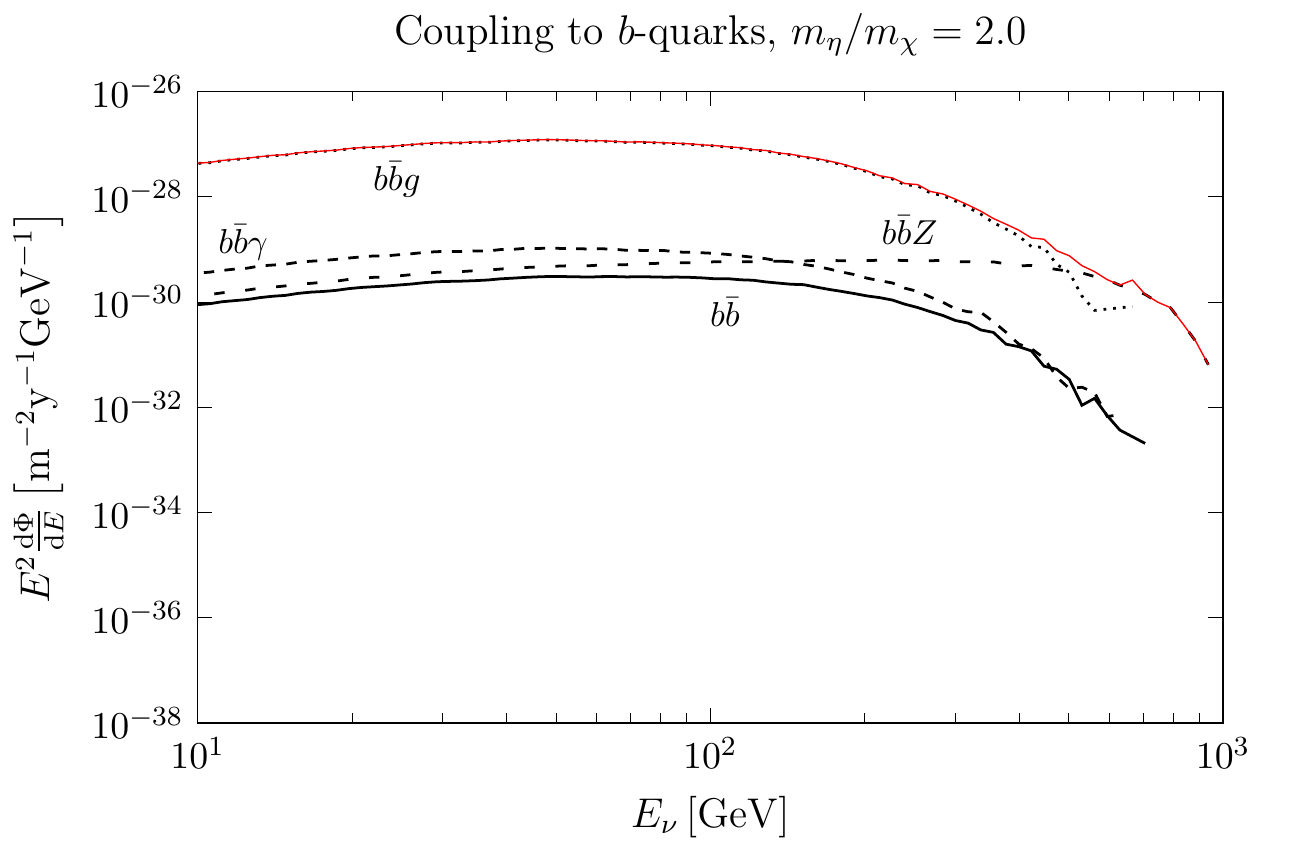}
\caption{ Sum of the differential $\nu_\mu$ and $\bar{\nu}_\mu$ flux at the Earth for $m_\chi = 1000 \GeV$  for coupling to up-quarks (upper panels) and bottom-quarks (lower panels) for the mass ratios $m_\eta / m_\chi =1.01$ (left panels) and 2.0 (right panels).}
\label{fig:spectra}
\end{center}
\end{figure}

We show in Fig.~\ref{fig:spectra} the differential flux at Earth of muon neutrinos and antineutrinos produced in the annihilation in the Sun of dark matter particles with mass $m_\chi=1\TeV$ which couple only to right-handed up-quarks (top panels) or to right-handed bottom-quarks (bottom panels), for the exemplary cases with mass degeneracy parameter $m_\eta/m_\chi=1.01$ (left panels) and  $m_\eta/m_\chi=2$ (right panels). In the scenario where the dark matter particle couples to up-quarks, the contribution to the neutrino flux from annihilations into $u\bar u$ is negligible, since this annihilation channel produces mostly light hadrons that, as explained above, do not contribute to the high energy neutrino flux and besides the branching fraction for this process is helicity suppressed. The annihilation into $u\bar u \gamma$ is not helicity suppressed and gives a larger contribution to the total flux, although still suppressed since this channel also produces mostly light hadrons. 
More relevant is then the annihilation into $u\bar u g$, which has the largest branching fraction (even when $m_\eta/m_\chi$ is sizable, {\it cf.} Fig. \ref{fig:branchingratios}) and moreover hadronizes producing a significant amount of heavy hadrons, which decay in flight contributing to the high energy neutrino flux. Finally, the annihilation channel into $u\bar u Z$ has, despite the fairly small branching fraction, a notable impact on the high energy neutrino flux, due to the hard neutrinos produced in the decay $Z\rightarrow \nu \bar \nu$. In fact, for $m_\chi=1000\GeV$, this channel is the dominant source of neutrinos when $E\gtrsim 200\GeV$. Therefore, this channel will be particularly important for IceCube, which is mostly sensitive to high energy neutrinos. 
Note also that in the limit $m_\chi\gg M_Z$, which is the relevant one for IceCube, the total neutrino spectrum is fairly insensitive to the degeneracy parameter, as long as $m_\eta/m_\chi\lesssim 2$. For small and moderate degeneracy parameters, the gauge bosons produced by the internal bremsstrahlung have similar energy spectra and, accordingly, also the neutrino specta. Furthermore, as discussed in \cite{Garny:2011ii}, when $m_\chi\gg M_Z$ the cross sections for the two-to-three processes are in fixed relations which depend only on the couplings of the gauge bosons to the final fermions. As a result, the total neutrino energy spectrum depends, for a given dark matter mass, very mildly on the degeneracy parameter. 

The conclusions for the case of a dark matter particle coupling to the up-quark also apply for the coupling to the bottom-quark. In this case, the annihilation cross section into $b\bar b$ is not as suppressed as into  $u\bar u$, due to the larger bottom quark mass, although for large dark matter masses the helicity suppression is still very strong and the dominant annihilation channels are the two-to-three processes. As apparent from the plot, also for the case of a dark matter particle coupling to the b-quark, the largest contribution to the neutrino spectrum at the highest energies is the annihilation $b \bar b Z$.

\section{Limits}
\label{sec:limits}

The muon (anti-)neutrino flux produced in the dark matter annihilations in the Sun generates a (anti-)muon signal which can be detected in a neutrino telescope. For a given neutrino spectrum we calculate the induced number of (anti-)muon events in IceCube following the approach of \cite{Scott:2012mq}, using the effective area presented in \cite{DanningerPhD}. The background consists of muons induced by the atmospheric neutrino flux and is given, together with the actual data, in \cite{Aartsen:2012kia}. For our analysis, it is necessary to choose a cut on the angle between the reconstructed muon direction and the Sun; to optimize the search we use, for each dark matter mass and mass ratio, the angle that gives the best constraint under a background only hypothesis. More details of our method of calculating the limits from the IceCube data can be found in appendix \ref{sec:StatisticalAnalysis}.

\begin{figure}
\begin{center}
\includegraphics[width=0.49\textwidth]{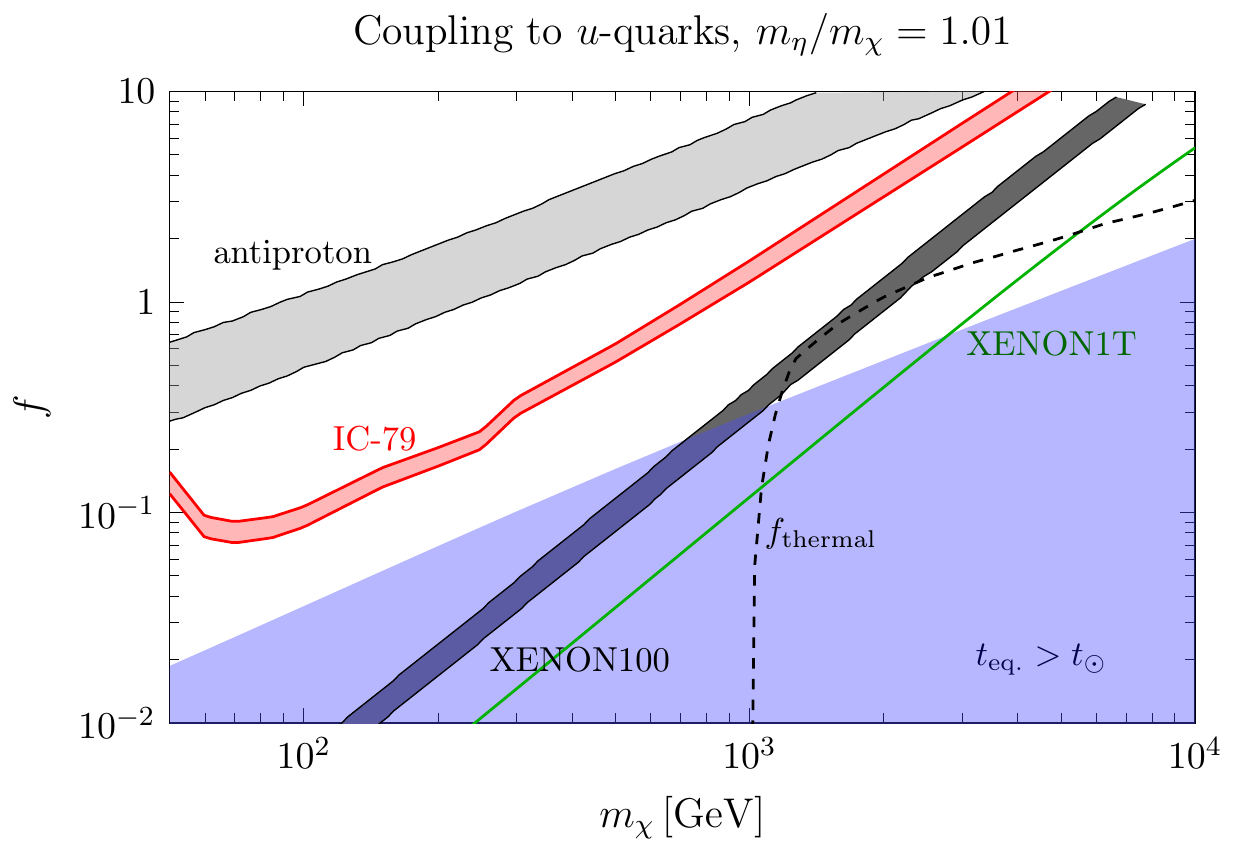}\\
\includegraphics[width=0.49\textwidth]{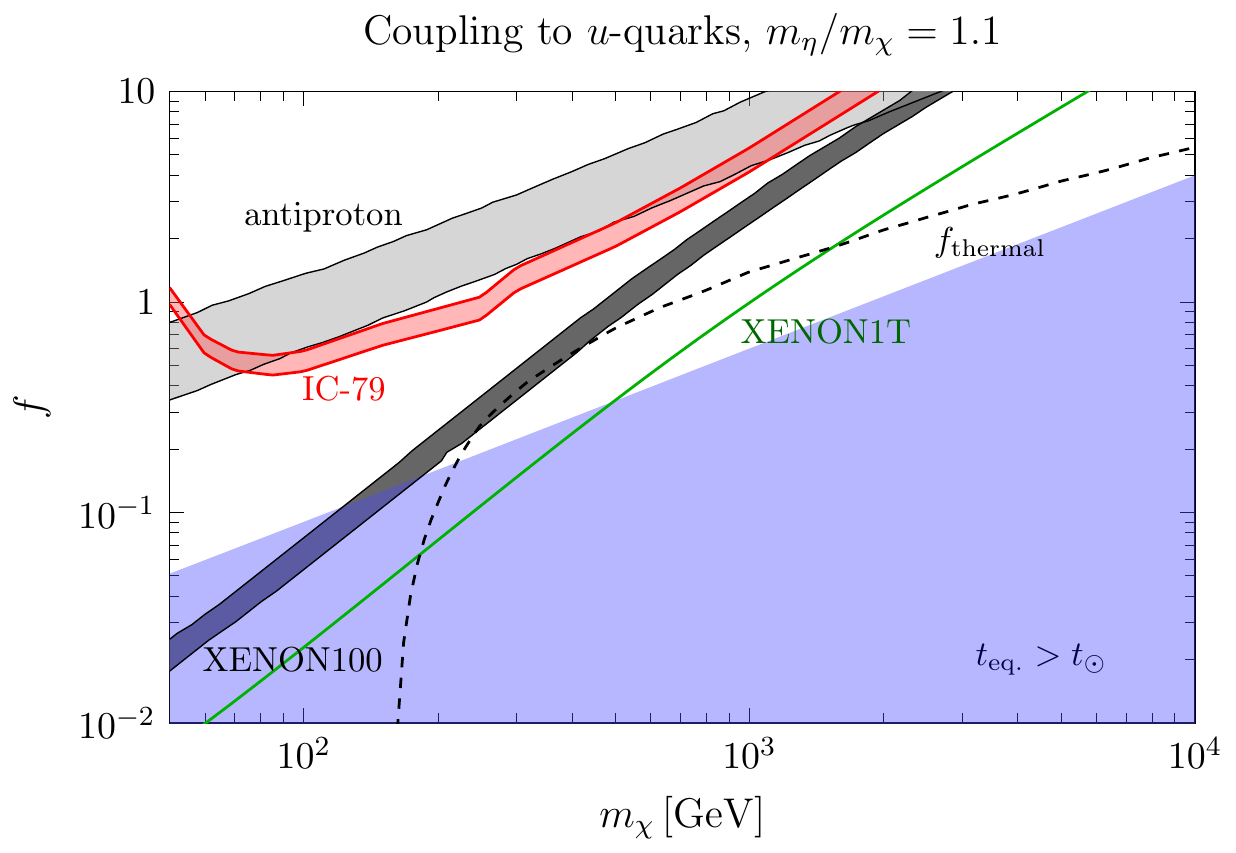} 
\includegraphics[width=0.49\textwidth]{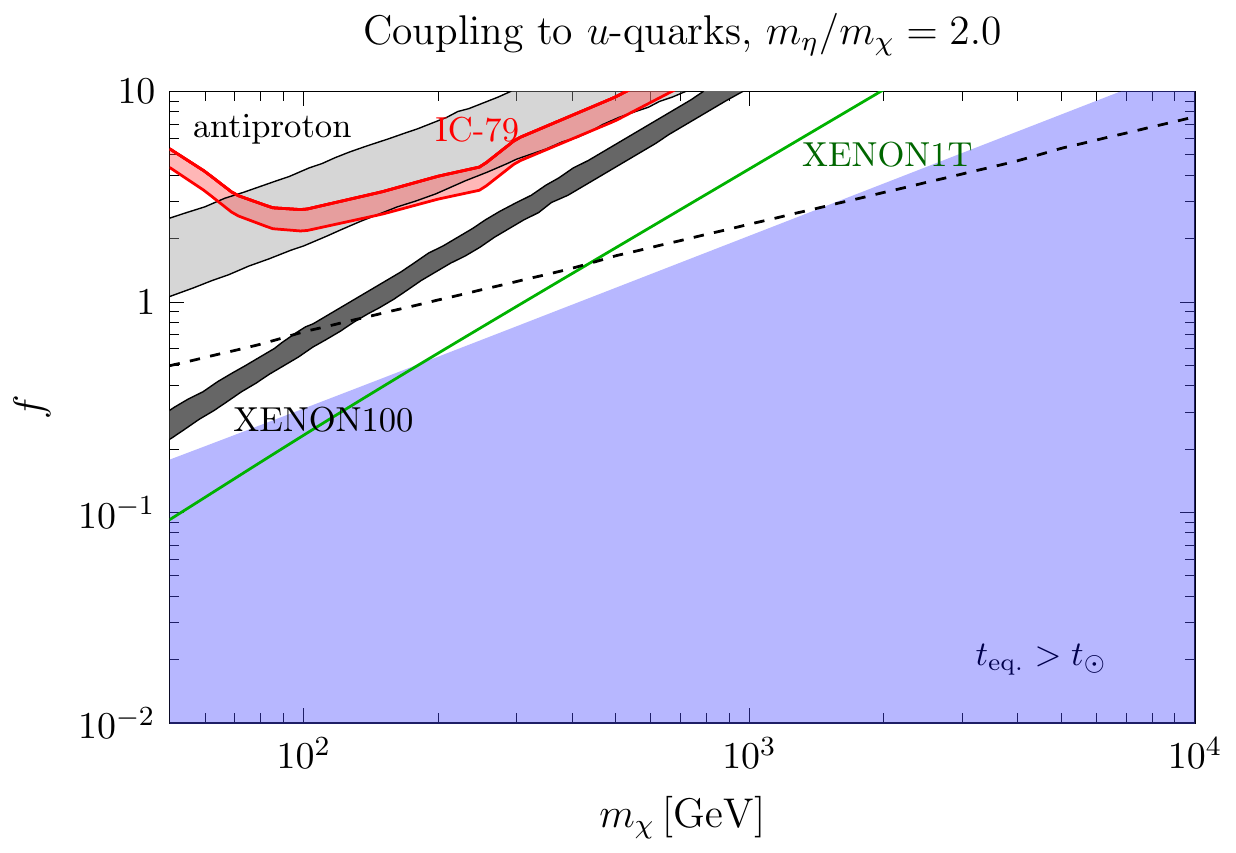}
\caption{Upper limits on the Yukawa coupling $f$ for coupling to up-quarks for $m_\eta/m_\chi=1.01$ (upper panel), 1.1 (lower left panel) and 2.0 (lower right panel) from IceCube-79 (light red), XENON100 (dark grey) and antiprotons (light grey). The width of the bands bracket the respective uncertainties. The shaded blue region shows the region of non-equilibration in the Sun while the dashed line shows the value of the coupling that generates the observed dark matter abundance via thermal production. We also show as a green line the projected limit from XENON1T.}
\label{fig:limits-upquark}
\end{center}
\end{figure}

The non-observation in IceCube-79 of an excess of events with respect to the expectations from the atmospheric background then allows to exclude regions of the parameter space of the model. We show in Fig.~\ref{fig:limits-upquark} as a light red band the limits on the coupling $f$ as a function of the dark matter mass, in a toy model where the dark matter particle only couples to the right-handed up-quarks for three exemplary choices of the degeneracy parameter, $m_\eta/m_\chi=1.01$, 1.1 and 2.0. The width of the band brackets the sensitivity of the upper limit to systematic uncertainties in the determination of the capture rate, which were estimated to be $3 \%$ ($25 \%$) for spin dependent (spin independent) capture \cite{DanningerPhD}, as well as to uncertainties in the determination of the astrophysical parameter $v_0$, the nuclear parameters $\Sigma_{\pi n}$, $\sigma_0$ and the second moments of the quark PDFs. The values of the corresponding uncertainties are chosen as in \cite{Garny:2012eb}. We also show for comparison the limits derived in \cite{Garny:2012eb} on the coupling $f$ from the non-observation of an excess in the cosmic antiproton-to-proton fraction measured by PAMELA (grey band), and which in this model stem mostly from the two-to-three process $\chi\chi\rightarrow q\bar q g$, or from the non-observation of a dark matter signal in the XENON100 experiment (dark blue band). Again, the width of the band brackets the sensitivity of the upper limit to the astrophysical and nuclear uncertainties. As apparent from the plot, IceCube-79 gives limits which are competitive, and for small degeneracies better, than the PAMELA antiproton limits. On the other hand, the IceCube-79 limits are worse than the XENON100 limits, due to the fact that this toy model has both spin dependent and spin independent interactions with matter, the latter being strongly constrained by direct search experiments. 

We also show in the plot the region of the parameter space where, for this toy model, captures and annihilations are in equilibrium in the Sun and therefore the annihilation rate reaches its maximum value. The shaded blue region then corresponds to model parameters where the neutrino signal from the Sun is very suppressed due to the inefficient dark matter capture and therefore observing an annihilation signal from the Sun becomes very challenging. As can be seen from the plot, IceCube-79, and especially XENON100, already probe a fairly large region of the parameter space where a signal of annihilations from the Sun could be found, especially for scenarios with large mass degeneracy. 
The XENON1T experiment will close in on the parameter space even further, concretely, the factor of $\sim 60$ increase in sensitivity projected in the mass range of interest for our analysis \cite{Aprile:2012zx}, will translate into an improvement in the limit on the coupling by a factor of $60^{1/4}\approx 2.8$. Note also that XENON100 rules out the values of the couplings necessary to achieve equilibriation inside the Earth, as can be checked from comparing Fig.~\ref{fig:equilibrium}, bottom plot, with Fig.~\ref{fig:limits-upquark}. Therefore, the observation of a high energy neutrino signal from the center of the Earth is very unlikely in this scenario.

\begin{figure}
\begin{center}
\includegraphics[width=0.49\textwidth]{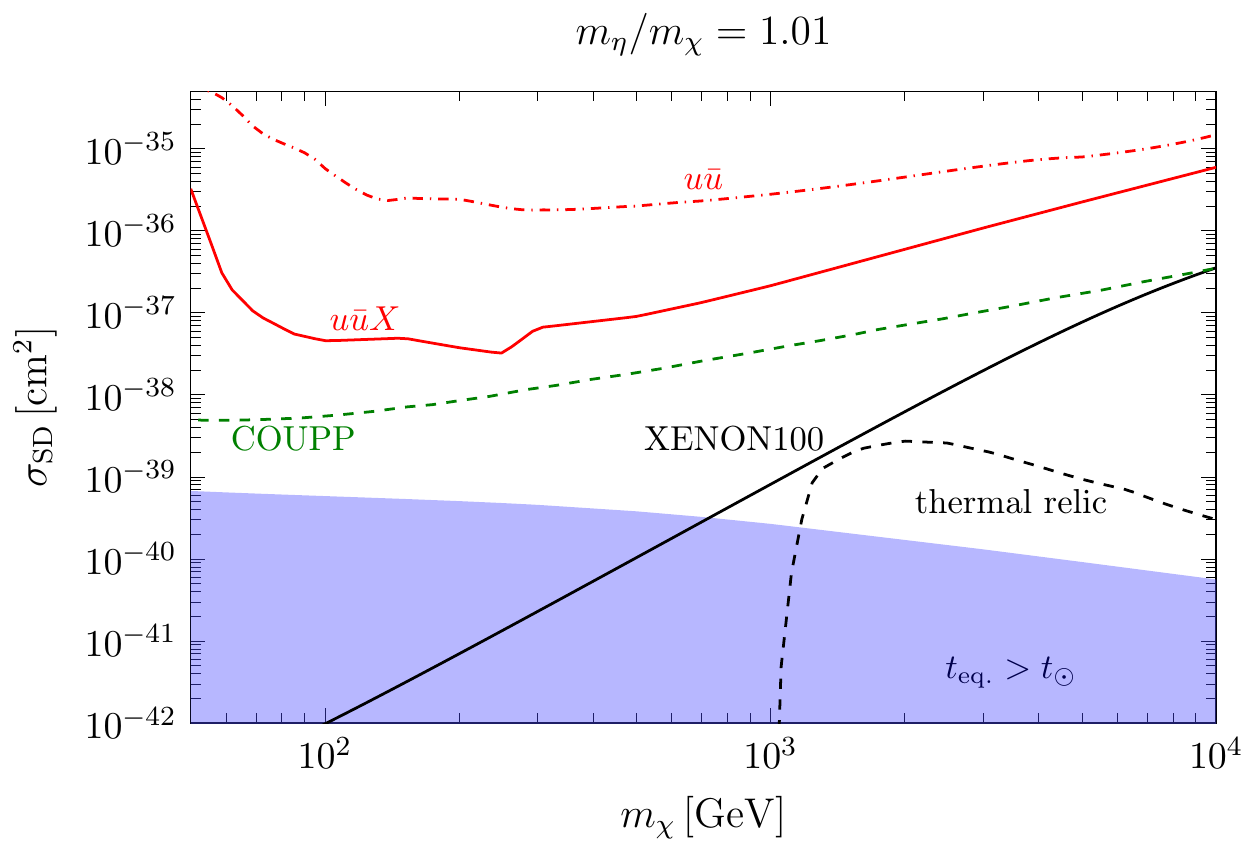} \\
\includegraphics[width=0.49\textwidth]{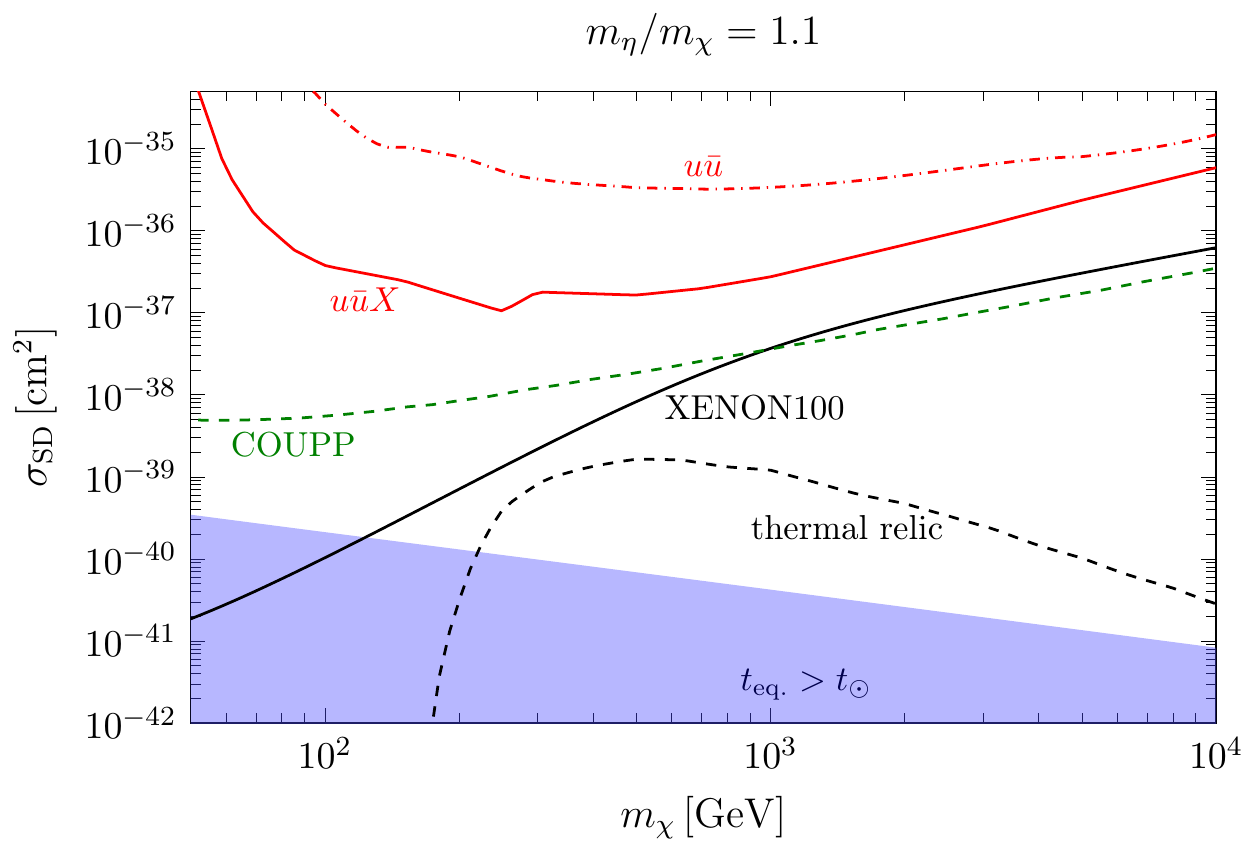} 
\includegraphics[width=0.49\textwidth]{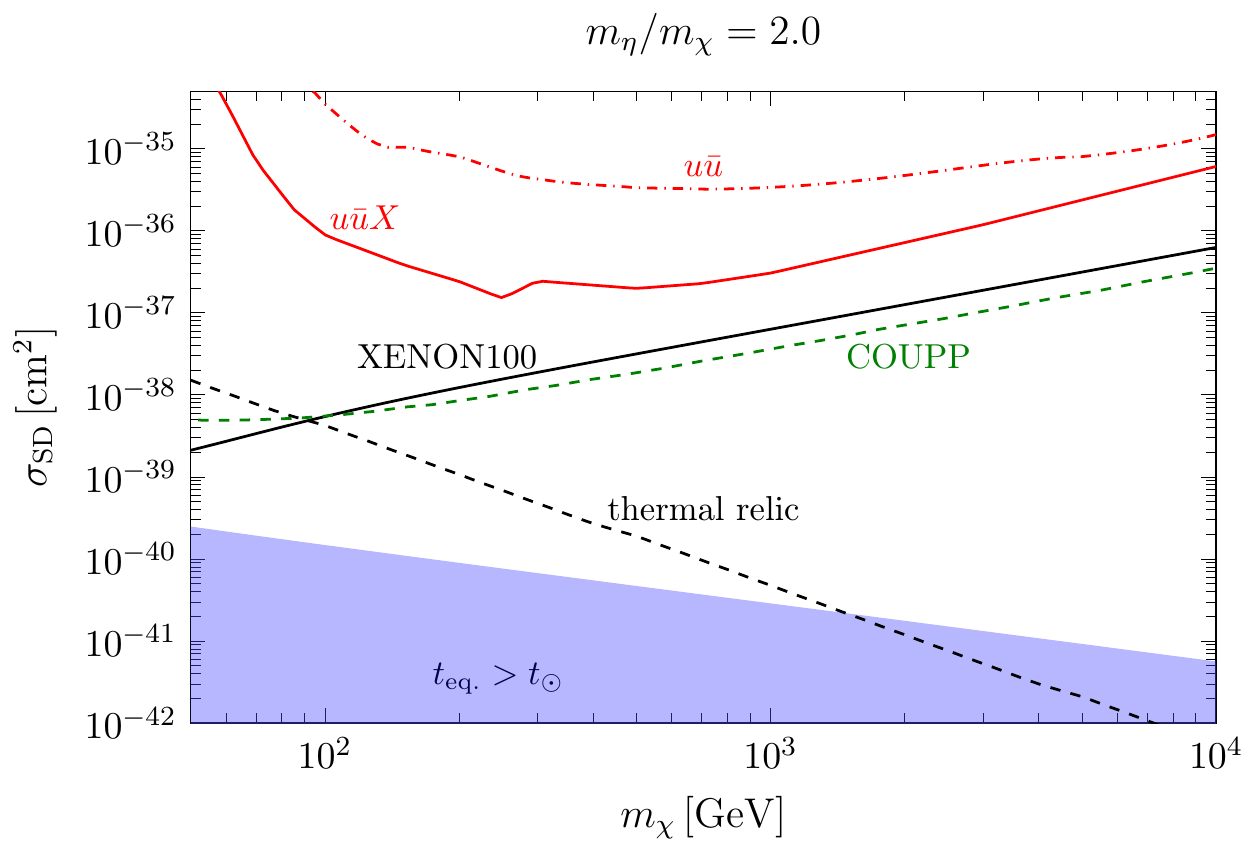}
\caption{Upper limit on the spin dependent interaction cross section for couplings to up-quarks for $m_\eta/m_\chi=1.01$ (upper panel), 1.1 (lower left panel) and 2.0 (lower right panel). The red dot-dashed line is the upper limit from IceCube-79 assuming only annihilations into $u\bar u$, while the solid line includes also the two-to-three annihilations into $u\bar{u} V$, with $V$ a gauge boson. The black solid line  is the upper limit on the spin dependent cross section of the model from XENON100 and the green dashed line, from COUPP. The shaded blue region shows the spin dependent cross section corresponding to the model parameters where equilibration between captures and annihilations does not occur in the Sun, while the dashed line, where the observed dark matter abundance can be generated via thermal production.}
\label{fig:limits-crosssection}
\end{center}
\end{figure}

The limits of the coupling can be straightforwardly translated into limits of the spin dependent cross section, as commonly presented in searches for neutrinos from dark matter annihilations in the Sun. The result is shown in Fig.~\ref{fig:limits-crosssection} for our exemplary cases $m_\eta/m_\chi=1.01$, 1.1 and 2.0. We show the limit on the spin dependent cross section that follows from assuming annihilations in the Sun only into $u \bar u$ as well as the limit that results from considering also the two-to-three annihilations. As apparent from the plot, including the three body annihilations in the Sun dramatically improves the limits on this scenario. We also show for completeness the experimental limits on the spin dependent cross section of this model from the COUPP \cite{Behnke:2012ys} and the XENON100 \cite{Aprile:2012nq} experiments. The lines always correspond, where applicable, to the choice of astrophysical and nuclear parameters giving the most conservative constraints. It should be stressed that the XENON100 limit in this plot refers to the limit on the spin dependent cross section from the non-observation of events in the XENON100 experiment {\it in this specific model}, which produces spin dependent and spin independent dark matter-nucleon interactions. The limits then differs from those reported in \cite{Garny:2012it,Aprile:2013doa}, where it was implicitly assumed a model with just spin dependent interactions.

\begin{figure}
\begin{center}
\includegraphics[width=0.49\textwidth]{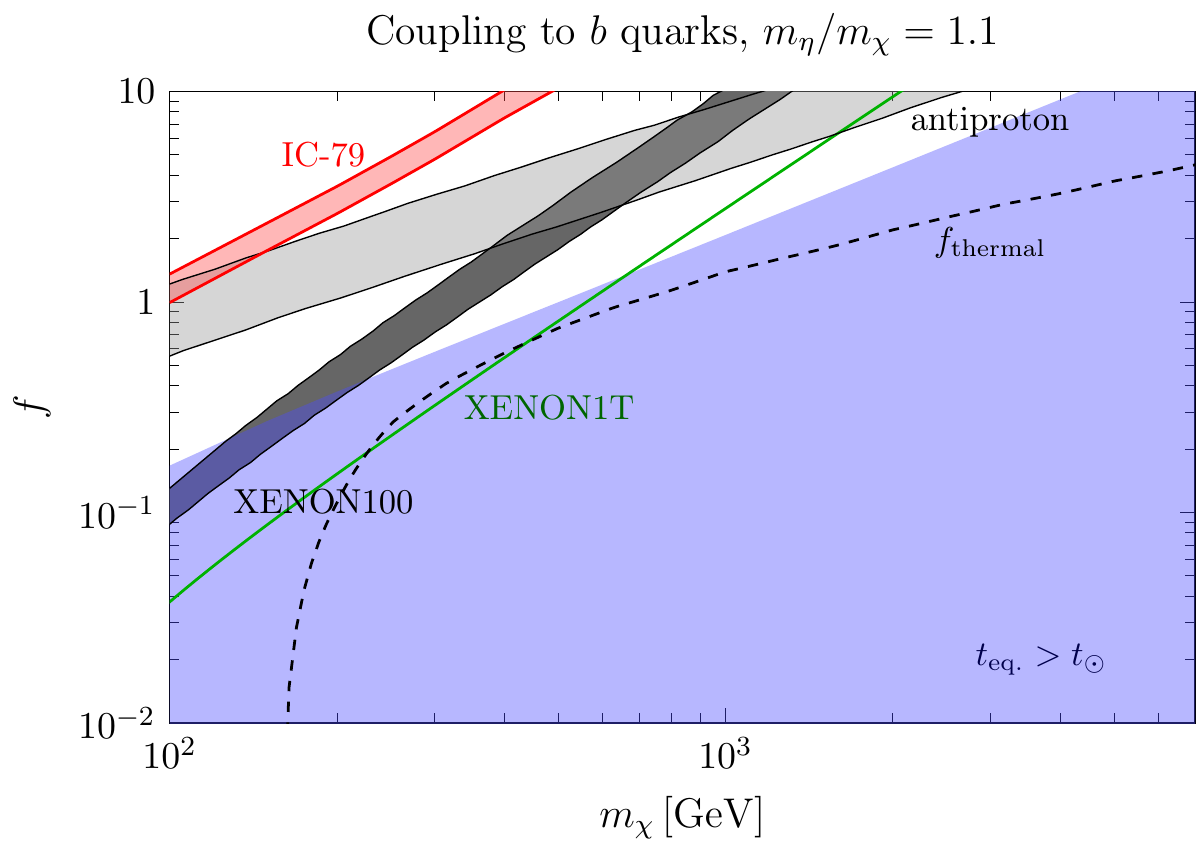}
\includegraphics[width=0.49\textwidth]{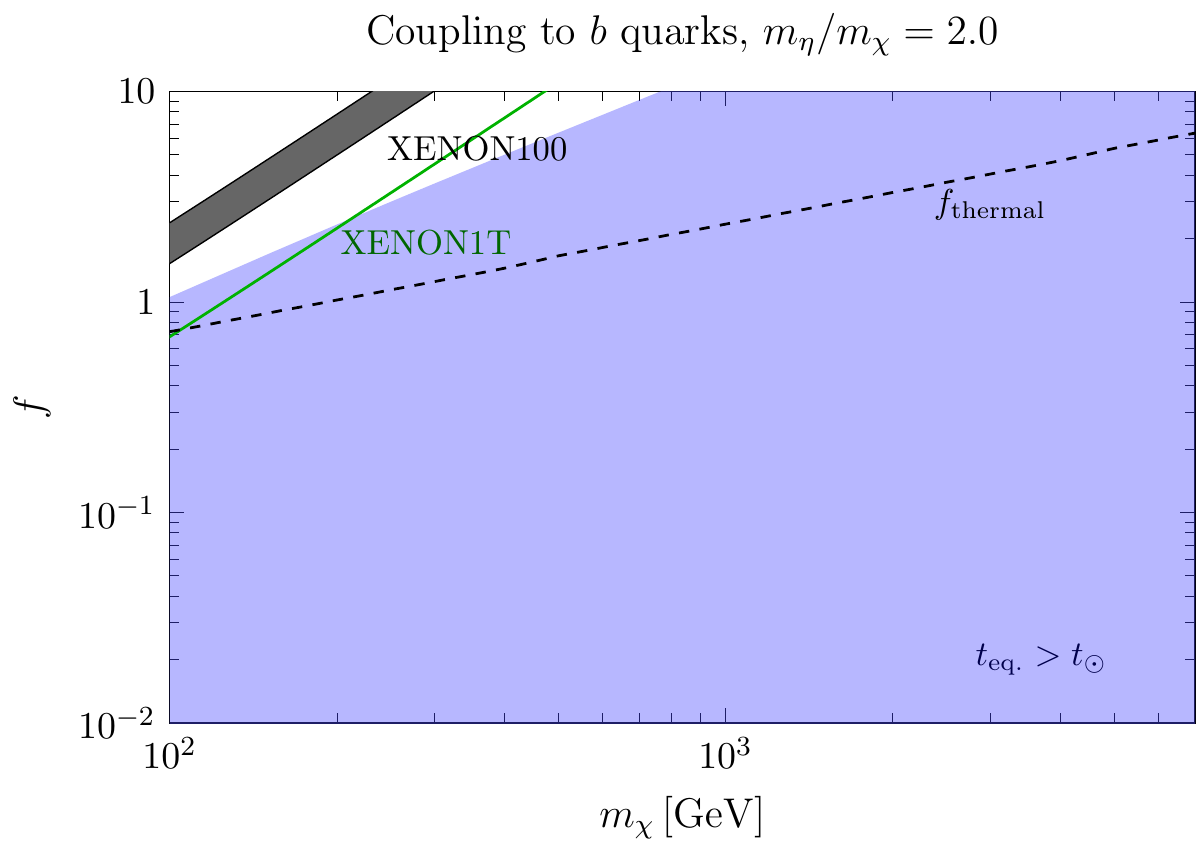}
\caption{ Same as Fig.~\ref{fig:limits-upquark} but for couplings to  bottom-quarks.}
\label{fig:limits-bquark}
\end{center}
\end{figure}

The limits for the scenario where the dark matter particle couples just to the right-handed bottom-quark are shown in Fig.~\ref{fig:limits-bquark}. As explained at the end of Section 2, we restrict our analysis to $m_\eta-m_\chi>2 m_q$, therefore we only show in this case the limits for $m_\eta/m_\chi=1.1$ and $m_\eta/m_\chi=2.0$, for which the previous condition is automatically fulfilled provided $m_\chi>100 \GeV$. Due to the suppressed spin dependent scattering rate, and as is apparent from the plots, for couplings to bottom-quarks the IceCube limits are rather weak, weaker than the antiproton limits and sensibly weaker than the XENON100 limits. It also follows from the plot that the XENON100 limits rule out already a large part of the parameter space where equilibration in the Sun is possible; the XENON1T experiment will improve the limit in the coupling $f$ by a factor $\sim 2.8$, if no signal is found.

In deriving these limits we have not made any assumption on how the dark matter particle was produced. It is then interesting to analyze the limits under the well motivated assumption that the dark matter particle was produced thermally in the early Universe. This fixes one of the three parameters of the toy model, $f$, $m_\chi$ or $m_\eta$. The value of the coupling constant required to produce the correct dark matter abundance is shown in Figs.~\ref{fig:limits-upquark},\ref{fig:limits-bquark} as a dashed line and lies well below the IceCube-79 limits, although close to, and for some choices of parameters above, the XENON100 limit. 

\begin{figure}
\begin{center}
\includegraphics[width=0.49\textwidth]{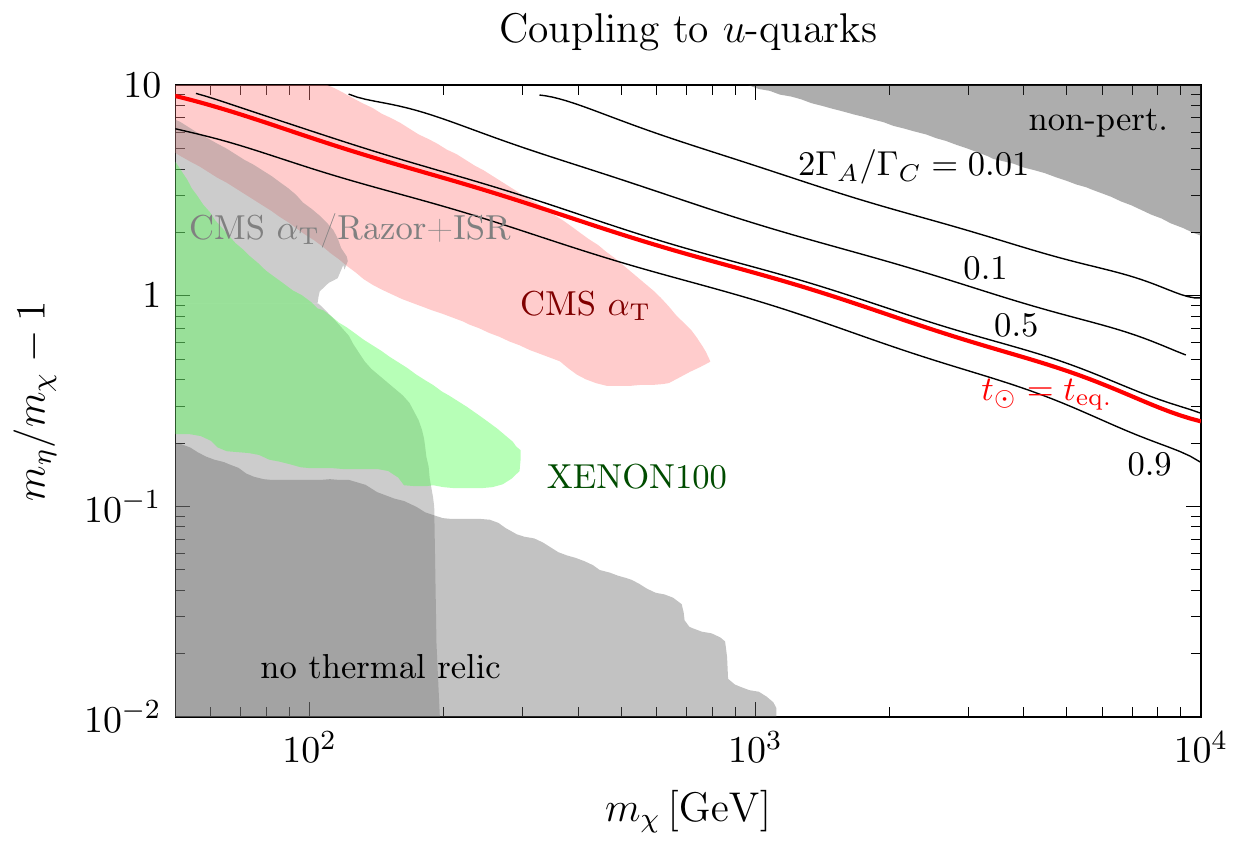}
\includegraphics[width=0.49\textwidth]{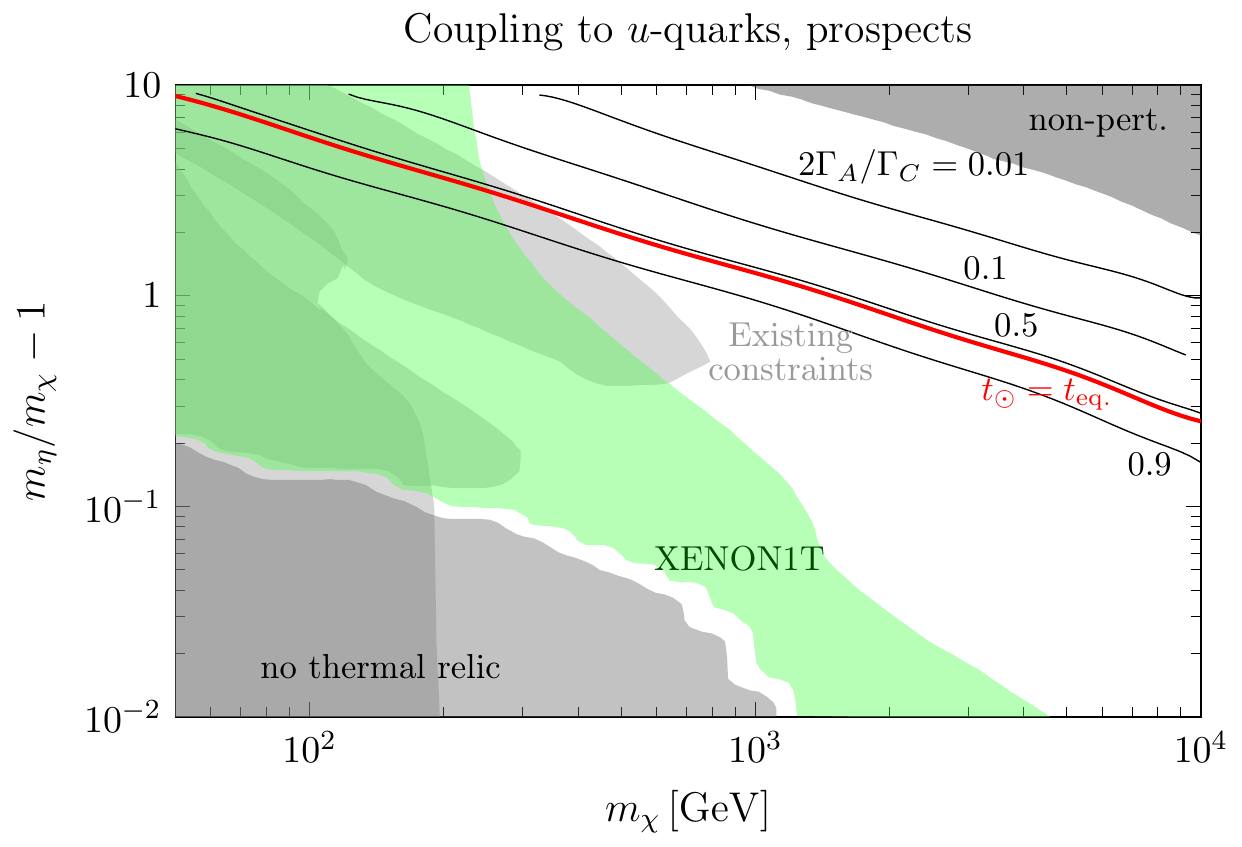}\\
\includegraphics[width=0.49\textwidth]{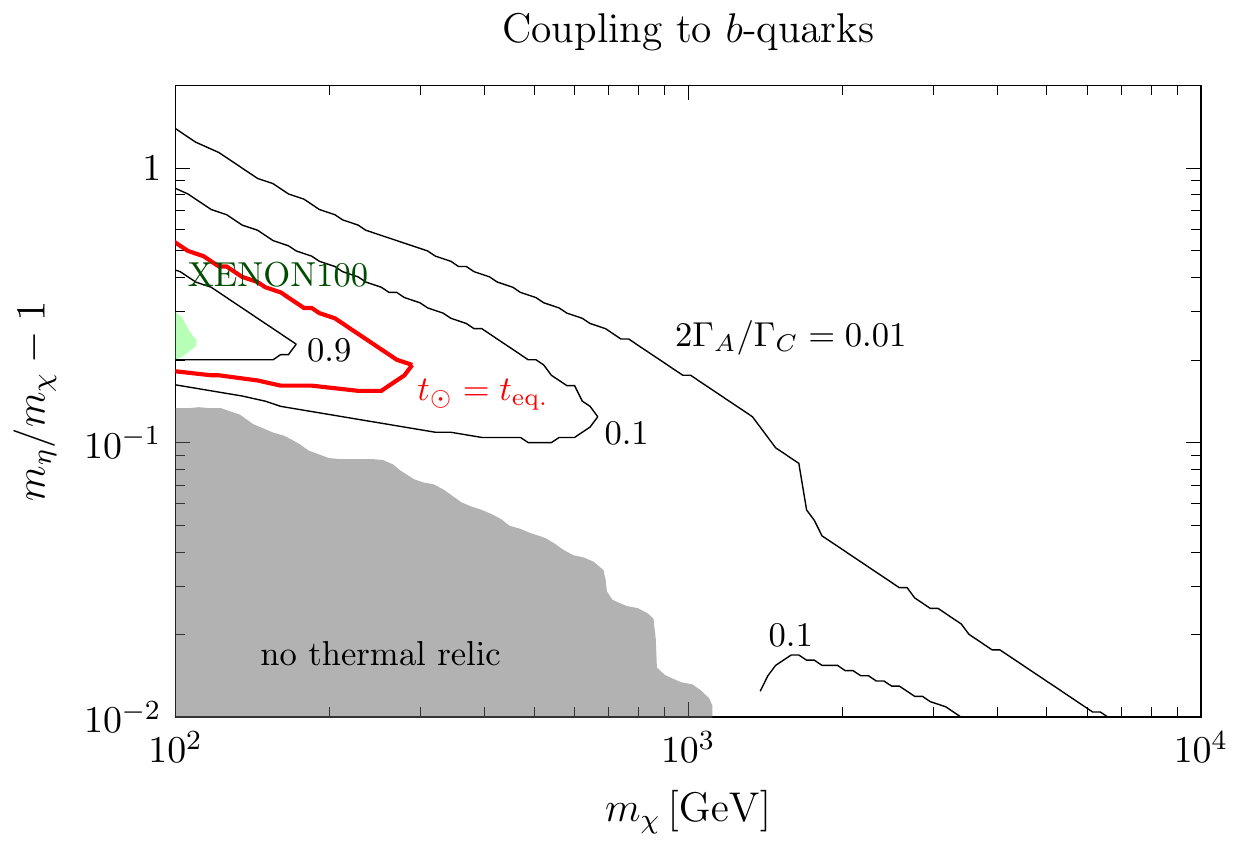}
\includegraphics[width=0.49\textwidth]{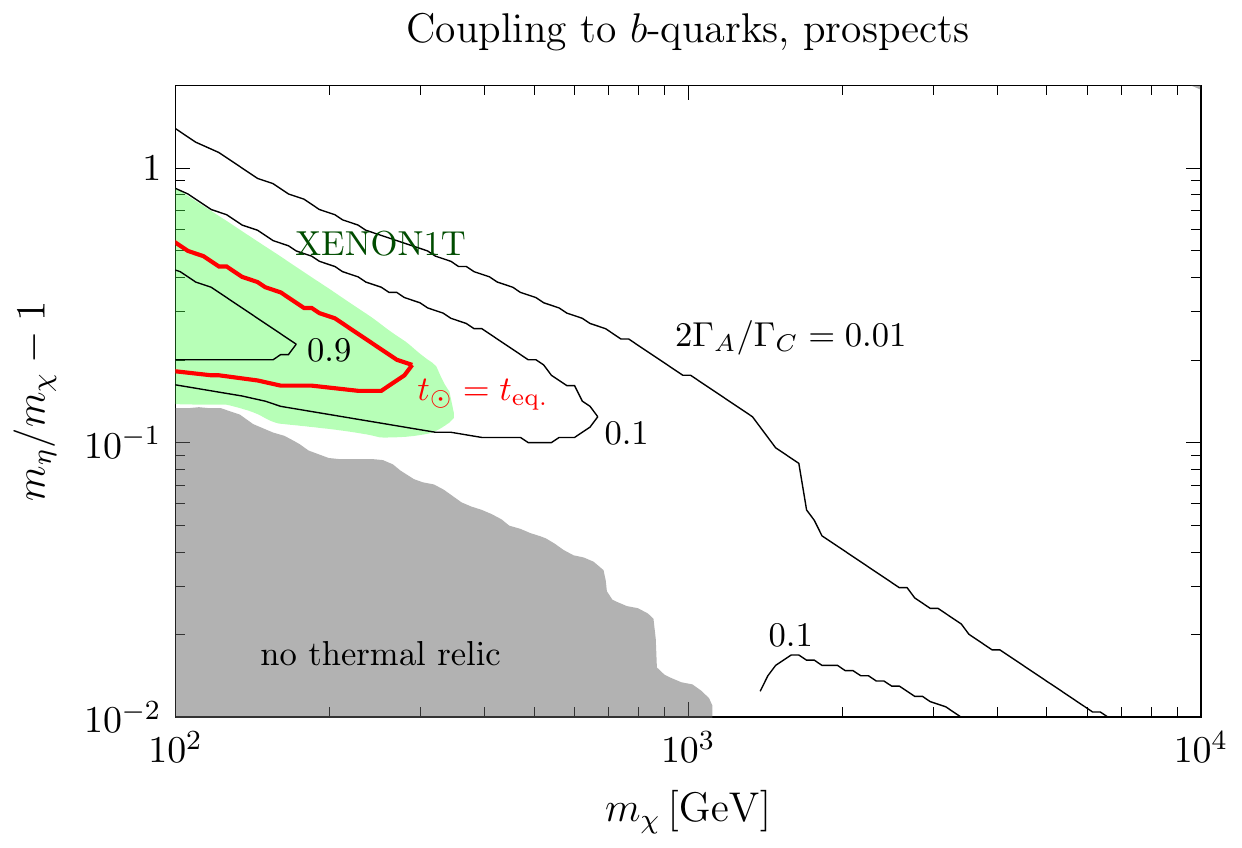}\\
\caption{Excluded regions of the parameter space of our toy model assuming couplings just to the up-quarks (upper panels) or to the bottom-quarks (lower panels) upon requiring that the observed dark matter abundance was thermally generated. The left panels show the present constraints from XENON100 (light green) and CMS (light red), as well as contour lines indicating the ratio of twice the annihilation rate over the capture rate in the Sun, which determines whether equilibration between these two processes is reached, {\it cf.} Eq.~(\ref{eq:SolutionDMDensitySun}). The thick red line indicates the parameters where the equilibration time equals the age of the Sun; above this line the high-energy neutrino flux from the Sun is strongly suppressed. The dark grey regions are theoretically not accessible, since they do not reproduce the observed relic abundance or because the Yukawa coupling becomes non-perturbative. The right panels show the projected sensitivity of XENON1T.}
\label{fig:limits-thermal}
\end{center}
\end{figure}

We further investigate in Fig. \ref{fig:limits-thermal} the complementarity of the various search strategies to probe this scenario, under the assumption that the observed dark matter abundance was thermally produced. We have calculated,  following \cite{Garny:2013ama} (see also \cite{An:2013xka,Bai:2013iqa,DiFranzo:2013vra}), the excluded regions of the parameter space, spanned by the dark matter mass and the mass splitting between $\eta$ and $\chi$, from the XENON100 results (light green) and from the LHC searches of a colored scalar by the CMS collaboration employing the $\alpha_T$-analysis based on 11.7 $\mathrm{fm^{-1}}$ at 8 TeV center of mass energy~\cite{Chatrchyan:2013lya} (light red). The present limits for the model with couplings to just right-handed up- (bottom-) quarks are shown in the upper (lower) left plots, together with the regions of the parameter space where it is impossible to generate thermally the observed relic abundance of dark matter particles (light gray) and the region where the perturbative calculation is not valid (dark gray).
We also show in the plot the lines of constant $2\Gamma_A/\Gamma_C$ (shown in black) and the line that marks the boundary of the parameter space where equilibration in the Sun is not attained, $t_{eq}<t_\odot$ (shown in red). Note that indirect search experiments with antiprotons and with high energy neutrinos are not yet sensitive to probe thermally produced dark matter particles, {\it cf.} Figs.\ref{fig:limits-upquark},\ref{fig:limits-bquark}, hence these experiments do not rule out any point in this parameter space. 
For couplings to up-quarks, XENON100 is the experiment that gives the strongest constraints on the parameter space at low mass splittings, and in particular rules out the possibility of observing a signal from the Sun for $m_\chi\lesssim 300\GeV$ when $m_\eta=1.2 \, m_\chi$; for larger mass splittings, it is the CMS experiment the one giving the strongest constraints. Future experiments, notably the XENON1T experiment, will cover a much larger region of the parameter space where a dark matter signal from the Sun could be observed, as apparent from Fig. \ref{fig:limits-thermal}, upper right plot.

For couplings to bottom-quarks, limits from present experiments are very weak, as can be seen in Fig. \ref{fig:limits-thermal}, lower left plot. In particular, the region of the parameter space where equilibration in the Sun takes place, around $m_\eta/m_\chi\simeq 1.1$ and $m_\chi\lesssim 300\GeV$, is barely constrained by the XENON100 experiment. Interestingly, the upcoming XENON1T experiment will be able to cover the whole region where equilibration is predicted to take place in the Sun, {\it cf.} Fig. \ref{fig:limits-thermal}, lower right plot. The non-observation of a dark matter signal at XENON1T will then have important implications for the feasibility of observing a high energy neutrino signal from the Sun in this concrete scenario.

\section{Conclusions}
\label{sec:conclusions}

The observation of an excess of high energy neutrinos in the direction of the Sun would be a clear signature of the annihilation of dark matter particles captured in the solar interior. In this paper we have investigated the prospects to observe this signature in a minimal scenario where the dark matter is a Majorana fermion that couples to a right-handed quark, concretely an up-quark or a bottom-quark, and a colored scalar via a Yukawa coupling. This scenario then allows to study the capture and the annihilation in a single Particle Physics framework. We have found that, for small and moderate mass splitting between the dark matter and the colored scalar, the most important channel in the calculation of the number of captured dark matter particles in the Sun is the two-to-three annihilation into $q \bar q g$. We have determined, for a given dark matter and colored scalar mass, the minimal value of the Yukawa coupling necessary to attain equilibrium between capture and annihilation. Upon comparing with the value of the coupling necessary to thermally produce the observed dark matter abundance, we conclude that for a thermal relic equilibration is never reached in the interior of Sun if the dark matter particle only couples to bottom-quarks. Equilibration in the interior of the Earth is not attained neither for couplings to up-quarks nor to bottom-quarks. Therefore, the observation of a high-energy neutrino flux from the Earth is, although not precluded in theory, very challenging in practice due to the very suppressed annihilation rate. 

We have also calculated the neutrino flux produced in the dark matter annihilations including the two-to-three annihilations into a quark-antiquark pair and a gauge boson. We have found that the neutrino flux at the Earth is, close to the kinematical endpoint, dominated by the annihilations into $q\bar q Z$, despite the rather small branching fraction in this channel. We have then derived an upper limit on the Yukawa coupling from the non-observation at IceCube of an excess of high-energy neutrino events in the direction of the Sun. For couplings to up-quarks, and small mass degeneracies between the dark matter particle and the colored scalar, the IceCube limits are stronger  than the limits on the same model parameters from the PAMELA measurements of the cosmic antiproton-to-proton fraction. 
On the other hand, for couplings to bottom-quarks the IceCube limits are weaker, due to the inefficient capture of dark matter particles in the Sun in this scenario. Both for couplings to up- and to bottom-quarks, the direct detection limits on the coupling from XENON100 are significantly stronger than the IceCube limits and already exclude regions of the parameter space where equilibration between capture and annihilation is reached inside the Sun. 

We have investigated in detail the limits on the parameter space under the assumption that the observed dark matter abundance was thermally generated, calculating the excluded regions from present searches of new physics at LHC and from XENON100. For couplings to bottom-quarks, the present limits are rather weak, while for couplings to up-quarks, they already exclude some regions of the parameter space. Future direct detection experiments, such as LUX and notably XENON1T will continue closing in on the the possibility of observing a high energy neutrino flux from the Sun. In particular, in the scenario where the dark matter couples just to right-handed bottom-quarks, XENON1T will cover the whole region of the parameter space where equilibration is reached in the solar interior. Therefore, the non-observation of a dark matter signal at XENON1T will then make very unlikely the possibility of observing a high energy neutrino signal from the Sun in this scenario.

\vspace{0.5cm}
\section*{Note Added}
After the completion of this work, the LUX experiment released new limits on the spin independent dark matter-nucleon scattering cross section \cite{Akerib:2013tjd}, which are approximately a factor of three better than the XENON100 limits in the range of masses of interest for our work. In view of the new results, our limits on the coupling constant $f$ are improved by approximately 30\%.

\vspace{0.5cm}
\section*{Acknowledgements}
We are grateful to  Sergio Palomares-Ruiz, Miguel Pato and Stefan Vogl for useful discussions. This work was partially supported by the DFG cluster of excellence ``Origin and Structure of the Universe,'' the Universit\"at Bayern e.V., the TUM Graduate School and the Studienstiftung des Deutschen Volkes.

\appendix
\section{Calculation of limits from the IceCube data}
\label{sec:StatisticalAnalysis}

In this appendix, we shortly review our method of converting a (anti-)neutrino flux from dark matter annihilations in the Sun into a number of (anti-)muon events in IceCube as well as the statistical method of extracting limits from the data, closely following \cite{Scott:2012mq}.

The number of expected muon signal events, $\sub{\theta}{S}$, for a given (anti-)neutrino flux $\frac{\mathrm{d} \sub{\Phi}{\nu/\bar{\nu}}}{\mathrm{d}E}$ is obtained by \cite{Scott:2012mq}
  \begin{align}
   \sub{\theta}{S} = \sub{t}{exp} \int\limits_{0}^{\infty}{L \left(E, \phi_{\text{cut}}\right) \left( A_\nu\left(E\right) \frac{\mathrm{d} \sub{\Phi}{\nu}}{\mathrm{d}E} + A_{\bar{\nu}}\left(E\right) \frac{\mathrm{d} \sub{\Phi}{\bar{\nu}}}{\mathrm{d}E} \right) \mathrm{d}E} \,.
   \label{eq:SignalNumberTelescope}
  \end{align} 
Here, $t_{\text{exp}}$ is the live-time of the 79-string analysis of IceCube given in \cite{DanningerPhD} and $A_{\nu / \bar{\nu}}\left(E\right)$ is the effective area for the detection of neutrinos/antineutrinos in IceCube, encapsulating all information about the efficiency of the IceCube detector as well as the scattering cross sections of neutrinos and the muon range in rock and ice.  We use the total effective area $A_{\nu}\left(E\right) + A_{\bar{\nu}}\left(E\right)$ for the IceCube 79-string configuration given in \cite{DanningerPhD} and we convert it into separate effective areas for neutrinos and antineutrinos using the method described in \cite{Arguelles:2012cf}. Besides, $L \left(E, \phi_{\text{cut}}\right)$ is the angular loss factor, defined as the probability that the angle of the reconstructed muon track, originated from an incoming neutrino with energy $E$, with respect to the direction of the Sun is smaller than a certain cut angle $\phi_{\text{cut}}$, namely
\begin{align}
L \left(E, \phi_{\text{cut}}\right) = 1 - \exp \left[ -\frac12 \left( \frac{\phi_{\text{cut}}}{\sigma_{\theta} \left( E \right)} \right)^2 \right] \,.
\end{align}
This assumes a 2D Gaussian point spread function of the muon track directions (see \cite{Scott:2012mq} for more details); we extract the corresponding median angle $\sigma_{\theta} \left( E \right)$ between the neutrino and the muon track from \cite{DanningerPhD}.

We then use the standard $CL_s$ method for calculating upper limits on $\sub{\theta}{S}$, based on a hypothesis test using the likelihood ratio
  \begin{align}
   X = \frac{\mathcal{L} \left(\sub{n}{obs} \mid \sub{\theta}{S} + \sub{\theta}{BG} \right) }{\mathcal{L} \left(\sub{n}{obs} \mid \sub{\theta}{BG} \right) } \,.
  \label{eq:likelihoodratio}
  \end{align}
For a definition of the likelihood functions and the corresponding $p$-values of the hypothesis test we refer to \cite{Scott:2012mq}. The number of observed events $n_{\text{obs}}$ and the number of off-source measured background events $\theta_{\text{BG}}$ as a function of the angle between the muon direction and the Sun are given in \cite{Aartsen:2012kia}.

For our analysis we choose the cut angle $\phi_{\text{cut}}$ such that the limit on the annihilation cross section is optimized. To this end, we use the number of neutrino events observed by IceCube in directions away from the Sun, which are presumably of atmospheric origin, and we make the plausible assumption that the background of atmospheric neutrinos in the direction of the Sun is identical to the background in these directions. We then calculate, for given $m_\chi$ and $m_\eta/m_\chi$, the limit on the annihilation cross section for eight different cut angles $\phi_{\text{cut}}$ between $3^{\circ}$ and $8.5^{\circ}$ (corresponding to the the bins in $\phi$ used in \cite{Aartsen:2012kia}) from the requirement that the dark matter signal would not produce an excess over the atmospheric neutrino background with a significance larger than 95\% C.L. In this step of the calculation, we use the off-source measured background events instead of the real data (i.e. the data in the direction of the sun); hence our procedure to determine the optimal cut angle is not biased by the actual measurements in the direction of the Sun, which might receive an exotic component.
Lastly, we select the cut angle that gives the best limit under this ``background-only'' hypothesis, and use it to calculate the limit on the annihilation cross section from the actual number of neutrino events measured by the IceCube collaboration in the direction of the Sun.

\begin{figure}
\begin{center}
\includegraphics[width=0.45\textwidth]{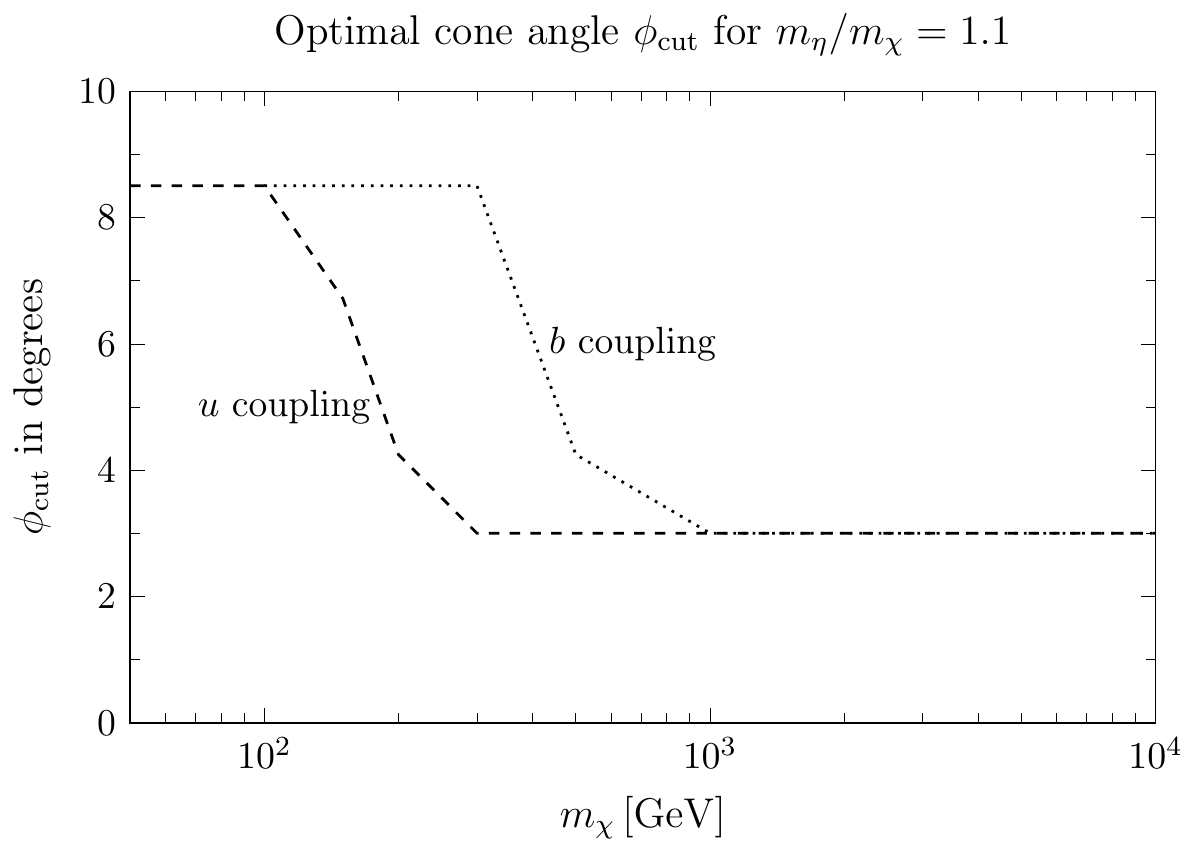}
\caption{Cut angle used in our analysis as a function of the dark matter mass for the scenario with couplings to up-quarks or bottom-quarks and mass ratio $m_\eta/m_\chi=1.1$. }
\label{fig:cut_angles}
\end{center}
\end{figure}

The optimal cut angle as a function of the dark matter mass is shown in Fig.~\ref{fig:cut_angles} for $m_\eta/m_\chi=1.1$ for couplings to up- and to bottom-quarks.  The differences in the spectra for couplings to up- or to bottom-quarks shown in Fig.~\ref{fig:spectra} get translated into different choices for the optimal cut angle in Fig.~\ref{fig:cut_angles}, due to the decrease of the average scattering angle between a neutrino and the muon produced in a charged-current interaction with the energy. Therefore, in a model with a neutrino spectrum peaked at high-energies, as in our toy model with coupling to up-quarks, small cut angles $\phi_{\text{cut}}$ are favored, as in this case most of the produced muons have directions within a small angle around the Sun; choosing a larger cut angle would increase the number of expected background events and hence lead to worse constraints. 
Conversely, in a model where the neutrino spectrum has an important soft component, as in our toy model with coupling to bottom-quarks, more signal events are found at larger angles and hence larger cut angles are favored, otherwise too many signal events would be cut, leading again to a worse constraint.

As a final comment, we note that the published 79-string data of IceCube \cite{Aartsen:2012kia} actually consists of three data sets, named ``Winter high'', ``Winter low'' and ``Summer low''. These are three non-overlapping data sets, each with its own distribution of $\sub{\theta}{S}$, $n_{\text{obs}}$ and $\theta_{\text{BG}}$ as well as its own effective area. The three event selections thereby correspond to different sets of experimental cuts, each optimized for a different signal model; in particular the  ``Summer low'' data set consists of downward going muons which were detected in DeepCore, using the surrounding IceCube strings as an active veto against atmospheric muons. 
While it would be in principle possible to use all three data sets in a combined likelihood for calculating the constraints, we choose the simpler (and conservative) strategy of calculating for each point in the parameter space three independent constraints using the three event selections and then choose the most constraining one. In our model, this turns out to be almost always the ``Winter high'' data set, except for dark matter masses below 200 GeV, where the ``Winter low'' or ``Summer low'' event selection can be relevant.



\bibliographystyle{JHEP-mod}
\bibliography{ITW}

\end{document}